\documentclass[useAMS,usenatbib]{mn2e}
\usepackage{amsmath}
\usepackage{amssymb}
\usepackage{amsfonts}
\usepackage{xcolor}
\usepackage{graphicx}
\usepackage{epstopdf}
\usepackage{enumerate}
\usepackage{subfigure}
\usepackage[hyperindex,breaklinks=true, colorlinks, citecolor=blue]{hyperref} 

\def\deg{\ifmmode^\circ\else$^\circ$\fi}

\def\ba{\begin{eqnarray}}
\def\ea{\end{eqnarray}}
\def\be{\begin{equation}}
\def\ee{\end{equation}}

\def\bdelta{{\delta }}

\makeatletter
\def\ref@jnl#1{{\rmfamily#1}}%
\newcommand\aj{\ref@jnl{AJ}}%
\newcommand\araa{\ref@jnl{ARA\&A}}%
\newcommand\apj{\ref@jnl{ApJ}}%
\newcommand\apjl{\ref@jnl{ApJ}}%
\newcommand\apjs{\ref@jnl{ApJS}}%
\newcommand\apss{\ref@jnl{Ap\&SS}}%
\newcommand\aap{\ref@jnl{A\&A}}%
\newcommand\aapr{\ref@jnl{A\&A~Rev.}}%
\newcommand\aaps{\ref@jnl{A\&AS}}%
\newcommand\baas{\ref@jnl{BAAS}}%
\newcommand\memras{\ref@jnl{MmRAS}}%
\newcommand\mnras{\ref@jnl{MNRAS}}%
\newcommand\pra{\ref@jnl{Phys.~Rev.~A}}%
\newcommand\prb{\ref@jnl{Phys.~Rev.~B}}%
\newcommand\prc{\ref@jnl{Phys.~Rev.~C}}%
\newcommand\prd{\ref@jnl{Phys.~Rev.~D}}%
\newcommand\pre{\ref@jnl{Phys.~Rev.~E}}%
\newcommand\prl{\ref@jnl{Phys.~Rev.~Lett.}}%
\newcommand\pasp{\ref@jnl{PASP}}%
\newcommand\pasj{\ref@jnl{PASJ}}%
\newcommand\ssr{\ref@jnl{Space~Sci.~Rev.}}%
\newcommand\nat{\ref@jnl{Nature}}%
\newcommand\iaucirc{\ref@jnl{IAU~Circ.}}%
\newcommand\aplett{\ref@jnl{Astrophys.~Lett.}}%
\newcommand\apspr{\ref@jnl{Astrophys.~Space~Phys.~Res.}}%
\newcommand\nphysa{\ref@jnl{Nucl.~Phys.~A}}%
\newcommand\physrep{\ref@jnl{Phys.~Rep.}}%
\newcommand\planss{\ref@jnl{Planet.~Space~Sci.}}%
\newcommand\procspie{\ref@jnl{Proc.~SPIE}}%

\makeatletter
\newcommand\footnoteref[1]{\protected@xdef\@thefnmark{\ref{#1}}\@footnotemark}
\makeatother

\title[Cosmology with HI intensity mapping]{Cosmological parameter forecasts for HI intensity mapping experiments using the angular power spectrum}
\author[L.\,C.\,Olivari]{L.\,C.~Olivari\thanks{\url{E-mail: lucas.olivari@postgrad.manchester.ac.uk}}$^1$, C.~Dickinson\thanks{\url{E-mail: clive.dickinson@manchester.ac.uk}}$^1$, R.\,A.~Battye$^1$, Y-Z.~Ma$^2$, A.\,A.~Costa,$^{3}$ \newauthor M.~Remazeilles$^1$, S.~Harper$^1$ \\
$^1$Jodrell Bank Centre for Astrophysics, Alan Turing Building, School of Physics \& Astronomy, The University of Manchester, \\
Oxford Road, Manchester, M13 9PL, U.K. \\ 
$^2$School of Chemistry and Physics, University of KwaZulu-Natal, Westville Campus, Private Bag X54001, Durban 4000, South Africa \\
$^3$Instituto de F\'{i}sica, Universidade de S\~{a}o Paulo, C.P. 66318, 05315-970, S\~{a}o Paulo, SP, Brazil
}
\begin{document}

\date{}

\setlength{\topmargin}{-15mm}

\pagerange{\pageref{firstpage}--\pageref{lastpage}} \pubyear{2002}

\maketitle

\label{firstpage}

\begin{abstract}

HI intensity mapping is a new observational technique to survey the large-scale structure of matter using the 21 cm emission line of atomic hydrogen (HI). In this work, we simulate BINGO (BAO from Integrated Neutral Gas Observations) and SKA (Square Kilometre Array) phase-1 dish array operating in auto-correlation mode. For the optimal case of BINGO with no foregrounds, the combination of the HI angular power spectra with \textit{Planck} results allows $w$ to be measured with a precision of $4\%$, while the combination of the BAO acoustic scale with \textit{Planck} gives a precision of $7\%$. We consider a number of potentially complicating effects, including foregrounds and redshift dependent bias, which increase the uncertainty on $w$ but not dramatically; in all cases the final uncertainty is found to be $\Delta w < 8\%$ for BINGO. For the combination of SKA-MID in auto-correlation mode with \textit{Planck}, we find that, in ideal conditions, $w$ can be measured with a precision of $4\%$ for the redshift range $0.35 < z < 3$ (i.e., for the bandwidth of $\Delta \nu = [350, 1050]$\,MHz) and $2\%$ for $0 < z < 0.49$ (i.e., $\Delta \nu = [950, 1421]$\,MHz). Extending the model to include the sum of neutrino masses yields a $95\%$ upper limit of $\sum m_\nu < 0.24$\,eV for BINGO and $\sum m_\nu < 0.08$\,eV for SKA phase 1, competitive with the current best constraints in the case of BINGO and significantly better than them in the case of SKA.

\end{abstract}


\section{Introduction}
\label{sec:intro}

In recent years, the cosmic microwave background (CMB) has been the main observational tool for cosmology. Sensitive measurements of the CMB power spectrum have been able to constrain the standard cosmological model with great accuracy \citep{Planck2016_1}. This standard model, known as $\Lambda$CDM, has only six parameters. More parameters lead to degeneracies and limit the constraining power of the CMB. This is due to the fact that the CMB is basically a snapshot of the Universe at $z \approx 1090$ and therefore gives us only 2-dimensional information about this particular time in the cosmic evolution. The next step towards precision cosmology will therefore need to use the extra information that can be obtained by measuring the large-scale structure of the Universe across cosmic time, principally at lower redshift, so that, in this case, the properties of dark energy can be constrained. This can be done through large optical or near-infrared galaxy surveys. Several of such surveys are now under way or in preparation, such as BOSS (SDSS-III) \citep{Dawson2013}, DES \citep{TDES2016}, eBOSS \citep{Zhao2016}, J-PAS \citep{Benitez2014}, DESI \citep{Levi2013}, LSST \citep{LSST2012}, and {\it Euclid} \citep{Amendola2013}.


It is possible, however, to perform redshift surveys in the radio waveband using the 21 cm radiation from neutral hydrogen (HI) to select galaxies \citep{Abdalla2005}.  One particular way to do a radio redshift survey is through a technique called HI intensity mapping (IM). HI IM measures the intensity of the redshifted 21\,cm line over the sky in a range of redshifts without resolving individual galaxies \citep{Madau1997ApJ, Battye2004, Peterson2006, Loeb2008}. The main advantages of HI IM compared to optical galaxy surveys are (i) a large volume of the Universe can be surveyed within a relatively short observing time; (ii) the redshift comes directly from the measurement of the redshifted 21\,cm line; (iii) all the signal is recorded, including gas in between galaxies; and (iv) HI is expected to be a good tracer of mass with minimal bias \citep[e.g.,][]{Padmanabhan2015, Penin2017}. Moreover, as cosmological measurements become more precise, independent probes and analyses will become important for understanding systematic errors \citep[e.g.,][]{Pourtsidou2016, Wolz2016_1, Hall2017, Carucci2017}.

Using the HI signal in cross-correlation with the WiggleZ galaxy survey data, the Green Bank Telescope (GBT) has made the first detection of the HI signal in emission at $z \approx 0.8$ \citep{Chang2010, Masui2013, Wolz2016}. This detection showed that the HI IM is indeed a feasible tool to study the large-scale structure of the Universe but also that it is extremely challenging. The main difficulties are the astrophysical contamination and the systematics that are present in the observed HI signal. At $\sim 1$\,GHz, the most relevant foregrounds are the Galactic emission, mostly synchrotron radiation, and the background emission of extragalactic point sources \citep{Battye2013}. These emissions are at least four orders of magnitude larger ($T \sim 10$\,K) than the HI signal ($T \sim 1$\,mK). The main systematics, for a single-dish experiment, are the $1/f$ noise, standing waves, and radio-frequency interference (RFI). The removal of these foregrounds and systematics is then a key challenge for the future HI IM experiments. Recent simulation work has shown that the existing foreground removal methods can recover the true HI power spectrum to within $5\%$ \citep{Shaw2014, Alonso2015, Wolz2014, Switzer2015, Olivari2016}, although over-subtraction of  the  HI  signal  can bias  the  recovered  spectrum  in  a  scale-dependent way. As we will discuss in this work, this may lead to a bias on some cosmological parameters if we want to use the shape of the power spectrum to constrain them. 

We consider two experimental setups. We focus on the BINGO (BAO from Integrated Neutral Gas Observations) \citep{Battye2013, Bigot-Sazy2015, Battye2016} concept, which is a proposed single-dish IM experiment that aims at mapping the HI emission at frequencies from 960 to 1260\,MHz ($z = 0.13$--0.48) over $\sim 3000$\,deg$^2$. BINGO corresponds to a ``stage II" experiment under the specifications of the Dark Energy Task Force \citep{Albrecht2006,Bull2015}. We also make a comparative study with the SKA (Square Kilometre Array) \citep{Maartens2015}. Here we consider the Phase 1 of the SKA-MID instrument, which will be composed of $\sim 200$ dishes working as a single-dish telescope. Although they will not be considered in this work, there are also other post-reionization epoch experiments planned such as the Canadian Hydrogen Intensity Mapping Experiment (CHIME) \citep{Bandura2014}, Tianlai Telescope \citep{Chen2012}, Five hundred metre Aperture Spherical Telescope (FAST) \citep{Smoot2017}, and Hydrogen Intensity and Real-Time Analysis experiment (HIRAX) \citep{Newburgh2016}.



There are two ways of analyzing the data obtained from HI IM experiments. First, we can use the 2-dimensional HI maps observed by them, one for each of their frequency channels (redshift bins), to calculate a set of angular power spectra. These power spectra can then be used to constrain the cosmological parameters exactly like it is done with the CMB angular power spectrum. No fiducial cosmology or filtering of the data is necessary in this case. Another way is to isolate the baryon acoustic oscillations (BAO) from the measured HI power spectrum and use the BAO signal alone to constrain the cosmological parameters, in special the dark energy equation-of-state $w$ \citep{Battye2013, Villaescusa2017}. In this technique, the power spectrum of the 2-dimensional distribution of the HI signal is computed and, assuming that the HI power spectrum is only  biased relative to the underlying dark matter distribution by some overall scale-independent constant, the acoustic scale, which is a function of the cosmological parameters, can be extracted with the help of some fitting formula of the BAO wiggles. Using the power spectrum amplitude and shape provides more information \citep{Rassat2008} and therefore allows a larger number of extensions of the $\Lambda$CDM model, such as models that have primordial fluctuations that are scale dependent or that include massive neutrinos, to be studied. A disadvantage is that the overall amplitude of the HI signal, governed by the $\Omega_{\mathrm{HI}}$ parameter, has to be taken into account in the analysis. This will then increase the uncertainty on $w$ and since the amplitude of the signal is important in this case, we become susceptible to any bias due to the foreground cleaning process.

The success of both techniques depends on a proper understanding of how the distribution of the HI is related to the distribution of the dark matter. To properly constrain our cosmological parameters we have then to assume an analytical expression for the HI bias and let the parameters of the assumed model to be free in our analysis \citep[e.g.,][]{Sarkar2016}. It has been stated before that in optical galaxy surveys the bias between the observed signal and the dark matter has a more significant effect on the full shape of the power spectrum than in the BAO wiggles \citep{Verde2002, Tegmark2004_1, Mehta2011}. In this work, however, we will find that the use of the power spectrum as our cosmological observable is more powerful than the use of the BAO wiggles alone, this being true even in the presence of extra nuisance parameters such as the HI bias and HI amplitude.

We organize this paper as follows. In Section~\ref{sec:experimental}, we describe the experimental parameters that are used in our BINGO and SKA-MID simulations. In Section \ref{sec:sims}, our simulations for the different components of the observed sky (HI signal, foregrounds and thermal noise) are described. In Section~\ref{sec:methods}, the component separation and statistical methods that are used in our analysis are described. In Section~\ref{sec:results}, we describe and discuss the results that we have obtained. Finally, in Section~\ref{sec:conclusion}, we make our final remarks and conclusions. 


\section{Experimental Setups}
\label{sec:experimental}

In this section, the two experimental setups that are used in this work, BINGO and SKA-MID phase 1, are described.

\begin{table*}
  \footnotesize
   \caption{Instrumental and observing parameters for the BINGO and SKA simulations.}
    \label{tab:instru}
    \begin{center}
      \begin{tabular}{|l|ccc}
        \hline
        Parameters & \multicolumn{3}{c}{Experiment} \\ \cline{2-4}
        \noalign{\smallskip}  
        &  BINGO & SKA-MID band 1 & SKA-MID band 2 \\ \hline \hline
        Redshift range, [$z_{\mathrm{min}}$, $z_{\mathrm{max}}$] & [0.13, 0.48] & [0.35, 3.0] & [0, 0.49] \\ \hline
        Bandwidth, [$\nu_{\mathrm{min}}$, $\nu_{\mathrm{max}}$] (MHz) & [960, 1260] & [350, 1050] & [950, 1421] \\ \hline
        Channel width, $\Delta \nu$ (MHz) & 7.5 & 17.5 & 11.25 \\ \hline
        Number of feed horns, $n_{\mathrm{f}}$ & 50  & 200 & 200 \\ \hline
        Sky coverage with the Galactic mask, $\Omega_{\mathrm{sur}}$ ($\mathrm{deg}^2$) & 2900 & 25000 & 25000 \\ \hline
        Observation time, $t_{\mathrm{obs}}$ (yrs) & 1 & 0.5 & 0.5 \\ \hline
        System temperature (mean), $T_{\mathrm{sys}}$ (K) & 50 & 31 & 17 \\ \hline
        Beamwidth (mean), $\theta_{\mathrm{FWHM}}$ (arcmin) & 40 & 122 & 66 \\ \hline
      \end{tabular}
    \end{center} 
\end{table*}

\subsection{BINGO}
\label{sec:bingo}

BINGO is a single-dish experiment that aims to map the HI emission at frequencies from 960 to 1260\,MHz ($z = 0.13$--0.48). We consider 40 frequency channels, which, as we choose the frequency channels to be equally spaced in the given bandwidth, gives us a channel width of 7.5\,MHz. The choice of the number of frequency channels is determined by the component separation performance, which, in general, improves with an increasing number of channels \citep{Olivari2016}. For this work, we find that 40 channels is enough to give us a good reconstruction of the HI signal.

Similar to \citet{Bigot-Sazy2015}, we consider circular Gaussian beams for the frequency channels of the experiment. For simplicity, we fix the full-width at half-maximum $\theta_{\mathrm{FWHM}}$ of the beam for all frequency channels to be equal to 40\,arcmin, which is the angular resolution of the BINGO telescope at 1\,GHz \citep{Battye2013}.

For our simulations, we assume that the BINGO telescope maps a $15^{\circ}$ declination strip centred at $-5^{\circ}$ as the sky drifts past the telescope. The declination of $-5^{\circ}$ has been chosen to minimize the foreground emission, which is lowest between $- 10^{\circ}$ and $10^{\circ}$ declination, and to maximize the survey area \citep{Dickinson2014}. For the observation time, we assume one full year of on-source integration.

We use a Galactic mask to cover the area of the sky where the synchrotron emission of our Galaxy is brightest. Our Galactic mask is given by one of the \textit{Planck} HFI masks (GAL070), which gives us a final sky coverage of $\approx 7\%$. To avoid boundary artifacts in our power spectrum estimation, we use a cosine apodization of width $3.5^{\circ}$ in our final mask. 

The instrumental parameters for the BINGO simulation are listed in Table~\ref{tab:instru}.

\subsection{SKA-MID}
\label{sec:ska}

Here we consider Phase 1 of the SKA-MID instrument. We assume there will be 200 15\,m diameter dishes with total-power dual polarization receivers, operating as a set of single-dish telescopes. This means that we are considering only the auto-correlation mode in this work, and not the cross-correlations i.e., visibilities from correlating the output from two dishes (interferometry) \citep{Bull2015,Santos2015}. We make this choice so that we consider only one type of IM experiment through this entire work: single-dish IM. The SKA-MID is going to have two bands: band 1, from 350\,MHz to 1050\,MHz ($0.35 < z < 3$), and band 2, from 950\,MHz to 1421\,MHz ($0 < z < 0.49)$. Band 1 has an expected average receiver temperature of $T_{\mathrm{rec}} = 23$\,K and band 2, $T_{\mathrm{rec}} = 15.5$\,K \citep{Bull2016}. The final system temperature can be approximated by
\begin{equation}
T_{\mathrm{sys}}(\nu) = T_{\mathrm{rec}} + 20 \, \mathrm{K} \, \left(\frac{408 \, \mathrm{MHz}}{\nu} \right)^{2.75}.
\end{equation}
The survey full-width at half-maximum is given by
\begin{equation}
\theta_{\mathrm{FWHM}} (\lambda) = 1.11 \, \frac{\lambda}{D},
\end{equation}
where $\lambda$ is the wavelength of the observed signal and $D$ is the SKA dish diameter ($D = 15$\,m). The scaling factor of 1.1 comes from measurements of the SKA-MID primary beam models (Robert Lehmensiek \textit{priv. comm.}). The reason for us to fix $T_{\mathrm{sys}}$ and $\theta_{\mathrm{FWHM}}$ for BINGO and not for SKA-MID is that the later's bandwidth is significantly larger than the former's, which makes the fixing of these two quantities to an average number to not be a good approximation for SKA-MID. The observing time that we consider is 6 months (around 4500 hours) and the survey area is $\Omega_{\mathrm{surv}} = 25000 \, \mathrm{deg}^2$  \citep{Bull2016}. As we do not consider the case of SKA with foregrounds, we do not use here any Galactic mask. The SKA dishes are assumed to be fixed at a constant elevation of $50^{\circ}$. Again, we will use 40 frequency channels, giving us a channel width of 17.5\,MHz for band 1 and 11.25\,MHz for band 2. In this work we do not consider the ``full" (phase 2) SKA, which is expected to be $\sim 10$ times more powerful than in phase 1 \citep{Maartens2015}.

The instrumental parameters for the SKA simulation are listed in Table~\ref{tab:instru}.

\section{Sky Model}
\label{sec:sims}

We now explain the components of our simulated radio sky: HI emission, HI shot noise, astrophysical foregrounds (Galactic synchrotron, Galactic free-free, and extragalactic point sources), and thermal noise from the BINGO and SKA experiments. We use the {\sc HEALPix} package \citep{Gorski2005} to produce maps with a resolution $N_{\mathrm{side}} = 128$ (pixel size $\approx 27$\,arcmin) and a maximum multipole $\ell_{\mathrm{max}} = 3 N_{\mathrm{side}} - 1 = 383$.

\subsection{HI Emission}
\label{sec:hi_emission}

The mean observed brightness temperature due to the average HI density in the Universe can be written as \citep{Battye2013, Olivari2016}


\begin{equation}
\label{backtemp_1}
\bar{T}(z) = 44 \, \mu \mathrm{K} \left( \frac{\Omega_{\mathrm{HI}}(z) h}{2.45 \times 10^{-4}} \right) \frac{(1 + z)^2}{E(z)},
\end{equation}
where $\Omega_{\mathrm{HI}}(z) = 8 \pi G \rho_{\mathrm{HI}}(z)/(3 H_0^2)$, with $h = H_0 / 100 \, \mathrm{km \, s}^{-1} \mathrm{Mpc}^{-1}$ and $E(z) = H(z) / H_0$.

In a linearly perturbed Universe, the 2D angular power spectrum of the HI intensity can be constructed over some frequency  range. Ignoring the effects of peculiar velocities (i.e., the redshift-space distortion effect) and the Sachs-Wolfe effect, we can obtain the 3D quantity $\delta T (r(z) \hat{n}, z)$ from Eq.~$\eqref{backtemp_1}$ by replacing $\rho_{\mathrm{HI}}$ (this quantity is present in the definition of $\Omega_{\mathrm{HI}}(z)$ above) with $\delta \rho_{\mathrm{HI}}$. Using a window  function $W(z)$,  which  we take  as  uniform  in the  observed  redshift  range, the projection on the sky of the temperature perturbation, $\delta T(\hat{n})$, is defined by \citep{Battye2013, Olivari2016}
\begin{align}
\delta T(\hat{n}) = \int \mathrm{d} z W(z) \bar{T}(z) \delta_{\mathrm{HI}}(r(z) \hat{n}, z), 
\end{align} where $\delta_{\mathrm{HI}} = \delta \rho_{\mathrm{HI}} / \rho_{\mathrm{HI}}$. 

Making a spherical harmonic transform, we obtain
\begin{align}
\delta T(\widehat{n}) = 4 \pi &\sum_{\ell, m} i^{\ell} \int \mathrm{d} z W(z) \bar{T}(z) \int \frac{\mathrm{d}^3 k}{(2 \pi)^3} \widehat{\delta}_{\mathrm{HI}} (\textbf{\textit{k}}, z) \nonumber \\  &\qquad {} \times j_{\ell} (k r(z)) Y_{\ell m}^*(\widehat{k}) Y_{\ell m} (\widehat{n}),
\end{align}
where $j_{\ell}(x)$ is the spherical Bessel function and $Y_{\ell m}(x)$ are the spherical harmonics. This expression gives the harmonic coefficients
\begin{align}
a_{\ell m} = 4 \pi i^{\ell} &\int \mathrm{d} z W(z) \bar{T}(z) \int \frac{\mathrm{d}^3 k }{(2 \pi)^3} \hat{\delta}_{\mathrm{HI}} (\textbf{k}, z) \\ \nonumber &\times j_{\ell} (k r(z)) Y_{\ell m}^* (\hat{k}).
\end{align} As the angular power spectrum $C_{\ell}$ is defined by the ensemble average 
\begin{equation}
C_{\ell} = \frac{1}{(2 \ell + 1)} \sum_m \vert a_{\ell m} \vert^2, 
\end{equation} 
the angular power spectrum for the HI signal is given by
\begin{align}
\label{cl_hi_full}
C_{\ell} &= \frac{2}{\pi} \int \mathrm{d} z W(z) \bar{T}(z) D(z) \int \mathrm{d} z' W(z') \bar{T}(z') D(z') \nonumber \\ &\times \int k^2 \mathrm{d} k \, b_{\mathrm{HI}}(z, k) b_{\mathrm{HI}}(z', k) P_c(k) j_{\ell} (k r(z)) j_{\ell} (k r(z')),
\end{align}
where $b_{\mathrm{HI}} (z, k)$ is the bias between the spatial distribution of the HI and the dark matter, $P_c(k)$ is the underlying dark matter power spectrum, and $D(z)$ is the growth factor for dark matter perturbations, which is defined such that $D(0) = 1$.

The calculation of the exact HI angular power spectrum via Eq.~\ref{cl_hi_full} is computationally demanding. We choose to simplify the calculation by using the Limber approximation \citep*{Limber1953}, which assumes a flat-sky and thin-shell model and is a good approximation to large $\ell$ ($\ell \gtrsim 50$) \citep{Loverde2008}, to perform the $k$ integral, 
\begin{align}
\left( \frac{2}{\pi} \right) \int k^2 \mathrm{d} k &b_{\mathrm{HI}}(z, k) b_{\mathrm{HI}}(z', k) P_c(k) j_{\ell} (k r) j_{\ell} (k r') \\ \nonumber &= b_{\mathrm{HI}}^2 \left(z, \frac{\ell + 1/2}{r} \right) P_c \left( \frac{\ell + 1/2}{r} \right) \frac{\delta (r - r')}{r^2}. 
\end{align} 
Writing $c \mathrm{d} z' = H_0 E(z') \mathrm{d} r'$, we can use the $\delta$-function to easily perform the $\mathrm{d} r'$ integral and deduce
\begin{align}
\label{c_ell_th}
C_{\ell} = \frac{H_0}{c} \int \mathrm{d} z E(z) &\left[ \frac{W(z) \bar{T}(z) D(z)}{r(z)} \right]^2  \nonumber \\ &\times b_{\mathrm{HI}}^2 \left(z, \frac{\ell + 1/2}{r} \right)  P_c \left( \frac{\ell + 1/2}{r} \right).
\end{align} 
The matter power spectrum $P_c (k)$ today can be computed using the software CAMB \citep{Lewis2000}. To generate our maps we use the routine {\sc synfast} \citep{Gorski2005}.

The main consequences of using the Limber approximation in our simulations are two: an overestimation of the auto-spectra and an underestimation of the cross-frequency spectra (the cross-frequency spectra are actually zero in the Limber approximation). These two effects, however, roughly compensate one another so that, in the end, the HI total power is correctly estimated. For example, if we define the total HI power by
\begin{equation}
S_{\mathrm{HI}} = \sum_{\ell} \sum_{i, j} C_{\ell}^{\mathrm{HI}}(z_i, z_j),
\end{equation} 
which is what is used to constrain the cosmological parameters in this work (see Eq. \ref{like}), we obtain, for a channel width of $\Delta \nu = 7.5$\,MHz and 40 frequency channels, which is the configuration that is assumed in our BINGO simulations, the following values

\begin{itemize}

\item[] Exact Calculation: $S_{\mathrm{HI}} = 0.0399\,\mathrm{mK}^2$,
\item[] Limber Approximation: $S_{\mathrm{HI}} = 0.0406\,\mathrm{mK}^2$.

\end{itemize} Thus, by assuming the Limber approximation, we obtain an error on the total HI power of just $1.7\%$, which justify the use of this approximation given the context of this work, i.e., the use of simulated emissions and the assumption that the redshift-space distortion effect is negligible.


\begin{figure*}
\centering
\subfigure{\includegraphics[width=0.33\textwidth]{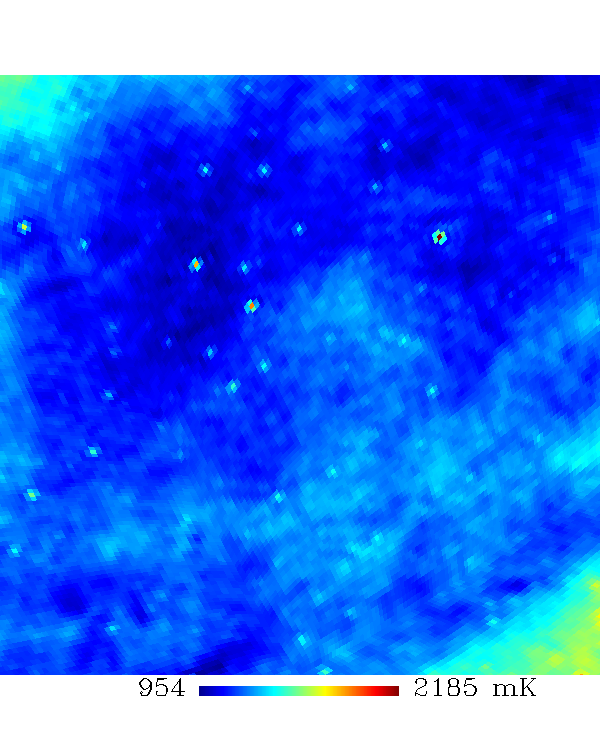}}
\subfigure{\includegraphics[width=0.33\textwidth]{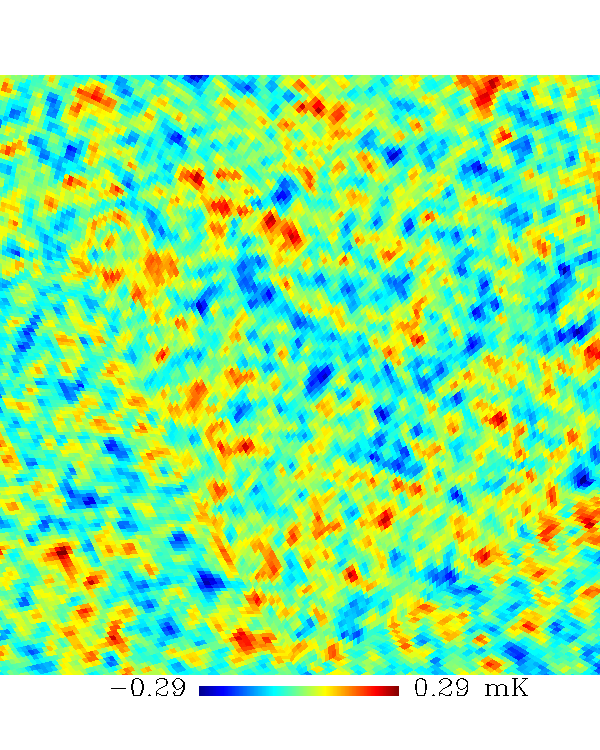}}
\caption{Simulated astrophysical emissions (Galactic synchrotron, Galactic free-free, and extragalactic point sources) (\textit{left}) and HI emission (\textit{right}) at 1$\,$GHz ($z \simeq 0.42$). The maps are centred at Galactic coordinates ($30^{\circ}, 120^{\circ}$) and cover a $60^{\circ} \times 60^{\circ}$ patch of the sky. The maps resolution is 40$\,$arcmin. The astrophysical foreground emissions are approximately four orders of magnitude brighter than the HI emission at this frequency (redshift). The colour scale is linear.}
\label{fig:signals}
\end{figure*}

\subsection{HI Shot Noise}
\label{sec:shoot}

Since the sources that emit the HI signal are discrete, the measured auto-spectra have a shot noise contribution as well as a clustering contribution, which we described in Section~\ref{sec:hi_emission}. Given an angular density of sources $\bar{N}(z)$, we have that the shot noise power spectrum is given by $C_{\ell}^{\mathrm{shot}} = \bar{T}^2(z) / \bar{N}(z)$ \citep{Hall2013}, where $\bar{T}(z)$ is the HI background temperature given by Eq.~\ref{backtemp_1}. By assuming a comoving number density of sources of $n_0 = 0.03 h^3 \, \mathrm{Mpc}^{-3}$ \citep{Masui2010}, we obtain an angular density of sources $\bar{N}(z)$ given by
\begin{equation}
\bar{N}(z) = \frac{n_0 c}{H_0} \int \frac{r^2(z)}{E(z)} \mathrm{d} z.
\end{equation} 


\subsection{Foregrounds}
\label{sec:foregrounds}

\subsubsection{Synchrotron Emission}

The synchrotron emission of our Galaxy arises from energetic charged particles accelerating in its magnetic field \citep{Rybicki2004}. The frequency scaling of synchrotron flux emission is often approximated in the form of a power-law, $I_{\nu} \propto \nu^{\alpha}$, over a limited range of $\nu$. In terms of the Rayleigh-Jeans brightness temperature, we have $T \propto \nu^{\beta}$, with $\beta = \alpha - 2$. There is evidence that the synchrotron spectral law is not a constant power-law, instead it has a curvature as the frequency increases \citep{Kogut2007}. In order to model this, we generalize the frequency dependency of the synchrotron brightness temperature to 
\begin{equation}
T \propto \nu^{\beta + C \log (\nu / \nu_p)},
\end{equation} 
where $C$ is  the  curvature  amplitude and $\nu_p$ is  a  pivot
frequency.  Positive  values  of $C$ flatten  and  negative ones steepen  the  spectral  law  for  increasing  frequency. Here we use $C = -0.3$ and $\nu_p = 1$\,GHz, which is a generalization of the 23\,GHz result of \citet{Kogut2012}. This makes the spectral response of the foregrounds more complex and more difficult to remove during component separation.

For  the  synchrotron  radiation  we  use  as a template the  reprocessed \citet{Haslam1982}  map  at  408\,MHz of \citet{Remazeilles2015}. As for the parameter $\beta$, we consider it to be spatially variable. For this, we use the model given by \citet{Miville2008}, which used WMAP intensity and polarization data to do a separation of the Galactic components. In this model the synchrotron spectral index has a mean value of $-3.0$ and a standard deviation of $0.06$.


\subsubsection{Free-Free Emission}

The free-free  emission  of our Galaxy arises from  the interaction  of  free  electrons  with  ions  in  its ionized  media \citep{Rybicki2004}. At radio frequencies, this comes from warm ionized gas with typical temperature of $T_{\mathrm{e}} = 10^4$\,K. The optical H$\alpha$ line is a good tracer of free-free emission although it requires corrections for dust absorption. To produce a free-free template we use dust-corrected H$\alpha$ map of \cite{Dickinson2003}, at an angular resolution of $1^{\circ}$. We have added small-scale fluctuations in order to increase the effective angular resolution required for our simulations. In doing this, we  follow  the  approach  used  in \citet{Delabrouille2013}, which simulates a Gaussian random field with the appropriated power spectrum and with the  same  statistics  (mean and variance) as the original template map and add this Gaussian signal to the original $1^{\circ}$ H$\alpha$ map.

The free-free brightness temperature is then given by
\begin{equation}
T_{\mathrm{ff}} \approx 10 \, \mathrm{mK} \, \left( \frac{T_e}{10^4 \, \mathrm{K}} \right)^{0.667} \left( \frac{\nu}{\mathrm{GHz}} \right)^{-2.1} \left( \frac{I_{\mathrm{H\alpha}}}{R} \right),
\end{equation}
where $I_{\mathrm{H\alpha}}$ is the H$\alpha$ template.



\subsubsection{Extragalactic Point Sources}

Extragalactic radio sources are an inhomogeneous mix of radio galaxies, quasars, star-forming galaxies, and other objects, which can be considered as point-like relative to the typical angular resolution used in a IM experiment. The contribution of point sources, $T_{\mathrm{ps}}$, to the observed sky can be calculated from the differential source count, $\mathrm{d} N/\mathrm{d} S$, representing the number of sources per steradian, $N$, per unit flux, $S$. A number of compilations of source counts are available. We choose to use data collected from continuum surveys at 1.4\,GHz between 1985 and 2009, as described in \citet{Battye2013}. We also use the fifth order polynomial that was fitted by \citet{Battye2013} to these data as our source count $\mathrm{d} N/\mathrm{d} S$.

Assuming that we can subtract all sources with flux density $S > S_{\mathrm{max}}$, where $S_{\mathrm{max}} = 10$\,Jy, we can calculate the mean temperature
\begin{equation}
\bar{T}_{\mathrm{ps}} = \left( \frac{\mathrm{d} B}{\mathrm{d} T} \right)^{-1} \int_0^{S_{\mathrm{max}}} S \frac{\mathrm{d} N}{\mathrm{d} S} \mathrm{d} S, 
\end{equation}
where $\mathrm{d} B/\mathrm{d} T = 2 k_B \nu^2/c^2$, $\nu$ is the observed frequency, $c$ is the speed of light and $k_B$ is the Boltzmann constant. We find $\bar{T}_{\mathrm{ps}} \approx 137$\,mK at 1.4\,GHz for $S_{\mathrm{max}} = 10$\,Jy.

There are two contributions to the fluctuations on this background temperature. The first is due to randomly (Poisson) distributed sources. The second, if the sources have a nontrivial two-point correlation function, is a contribution due to clustering.


Poisson distributed sources have, for $S_{\mathrm{max}} \lesssim 0.01$\,Jy, a white power spectrum given by
\begin{equation}
C^{\mathrm{Poisson}}_{\ell} = \left( \frac{\mathrm{d} B}{\mathrm{d} T} \right)^{-2} \int_0^{S_{\mathrm{max}}} S^2 \frac{\mathrm{d} N}{\mathrm{d} S} \mathrm{d} S.
\end{equation}
In the limit of a large number of sources, the intensity distribution becomes well approximated by a Gaussian distribution. However, for $0.01 < S< 10$\,Jy, the source density on the sky becomes too low and we must inject these bright sources in real (map) space. To do this, we first estimate the mean brightness temperature by
\begin{equation}
T_{\mathrm{ps}}(\nu, \hat{n}) = \left( \frac{\mathrm{d} B}{\mathrm{d} T} \right)^{-1} \Omega_{\mathrm{pix}}^{-1} \sum_{i = 1}^{N} S_i(\nu), 
\end{equation}
where $S_i(\nu)$ is the flux of the point source $i$ at frequency $\nu$ and $\Omega_{\mathrm{pix}}$ is the pixel size. We then randomly distribute in the sky $N$ of sources with flux $S(\nu)$ such that these sources respect our source count.

The power spectrum due to the clustered sources can be simply estimated as $C_{\ell}^{\mathrm{cluster}} = w_{\ell} \bar{T}_{\mathrm{ps}}^2$, where $w_{\ell}$ is the Legendre transform of their angular correlation function, $w(\theta)$. The clustering of radio sources at low flux densities ($<$10\,mJy) is not well known. To make an estimate, we use $w(\theta)$ measured from NVSS, which can be approximated as $w(\theta)\approx (1.0 \pm 0.2) \times 10^{-3} \theta^{-0.8}$ \citep{Overzier2003}. Legendre transforming this using a numerical calculation gives $w_{\ell} \approx 1.8 \times 10^{-4} \ell^{-1.2}$.

We assume a power-law frequency scaling for the point source brightness temperature, $T_b \propto \nu^{\alpha}$, and randomly choose the value of $\alpha$ for each pixel of the simulated map from a Gaussian distribution,
\begin{equation}
P(\alpha) = \frac{1}{\sqrt{2 \pi \sigma^2}} \exp \left[- \frac{(\alpha - \alpha_0)^2}{2 \sigma^2} \right],
\end{equation}
with a mean of $\alpha_0 = -2.7$ and a width distribution of $\sigma = 0.2$ \citep{Bigot-Sazy2015}.

The astrophysical emissions are much brighter than the HI emission. In Fig \ref{fig:signals}, we show a patch of the foreground sky at 1$\,$GHz and a patch of the HI fluctuations at the same frequency. We choose an area of the sky outside the Galactic plane, where the Galactic emission is brightest, because we use a Galactic mask when performing the component separation step in our analysis. Nevertheless, as can be seen in this figure, the difference in scale between the foregrounds and the HI signal is significant even in the areas of the sky where the Galactic emissions are minimal.


\subsection{Thermal Noise}
\label{sec:t_noise}

Similar to \citet{Olivari2016}, we consider only thermal noise in this work. This means that the simulated noise respects a uniform Gaussian distribution over the sky. Consequently, we do not consider $1/f$ noise, which has been considered in \citet{Bigot-Sazy2015}, for instance, and other sources of systematic errors, such as standing waves, calibration errors and RFI, which are going to be present in the real data. The theoretical noise level per beam of a total-power, dual polarization, single-dish experiment is given by the radiometer equation \citep{Wilson_book}
\begin{equation}
\label{thermal_noise}
\sigma_{\mathrm{t}} = \frac{T_{\mathrm{sys}}}{\sqrt{t_{\mathrm{pix}} \Delta \nu}},
\end{equation}
where $\Delta \nu$ is the frequency channel width, $T_{\mathrm{sys}}$ is the system temperature, and $t_{\mathrm{pix}}$ is the integration time per beam defined by
\begin{equation}
t_{\mathrm{pix}} = n_{\mathrm{f}} t_{\mathrm{obs}} \frac{\Omega_{\mathrm{pix}}}{\Omega_{\mathrm{sur}}},
\end{equation}
where $n_{\mathrm{f}}$ denotes the number of feed horns, $t_{\mathrm{obs}}$ is the total integration time, $\Omega_{\mathrm{sur}}$ is the survey area, and $\Omega_{\mathrm{pix}} = \theta_{\mathrm{FWHM}}^2$ is the beam area. 

With the help of Eq.~\eqref{thermal_noise}, we can find the noise amplitude of BINGO and SKA-MID. Assuming $\Delta \nu = 7.5$\,MHz, the BINGO thermal noise sensitivity per beam and per channel is $\sigma_{\mathrm{t}} = 26$\,$\mu$K. The SKA-MID band 1 average thermal noise sensitivity per beam is $\sigma_{\mathrm{t}} = 7.5$\,$\mu$K and that the SKA-MID band 2 average thermal noise sensitivity per beam is $\sigma_{\mathrm{t}} = 9.1$\,$\mu$K. 

\begin{figure*}
\centering
\subfigure{\includegraphics[width=0.49\textwidth]{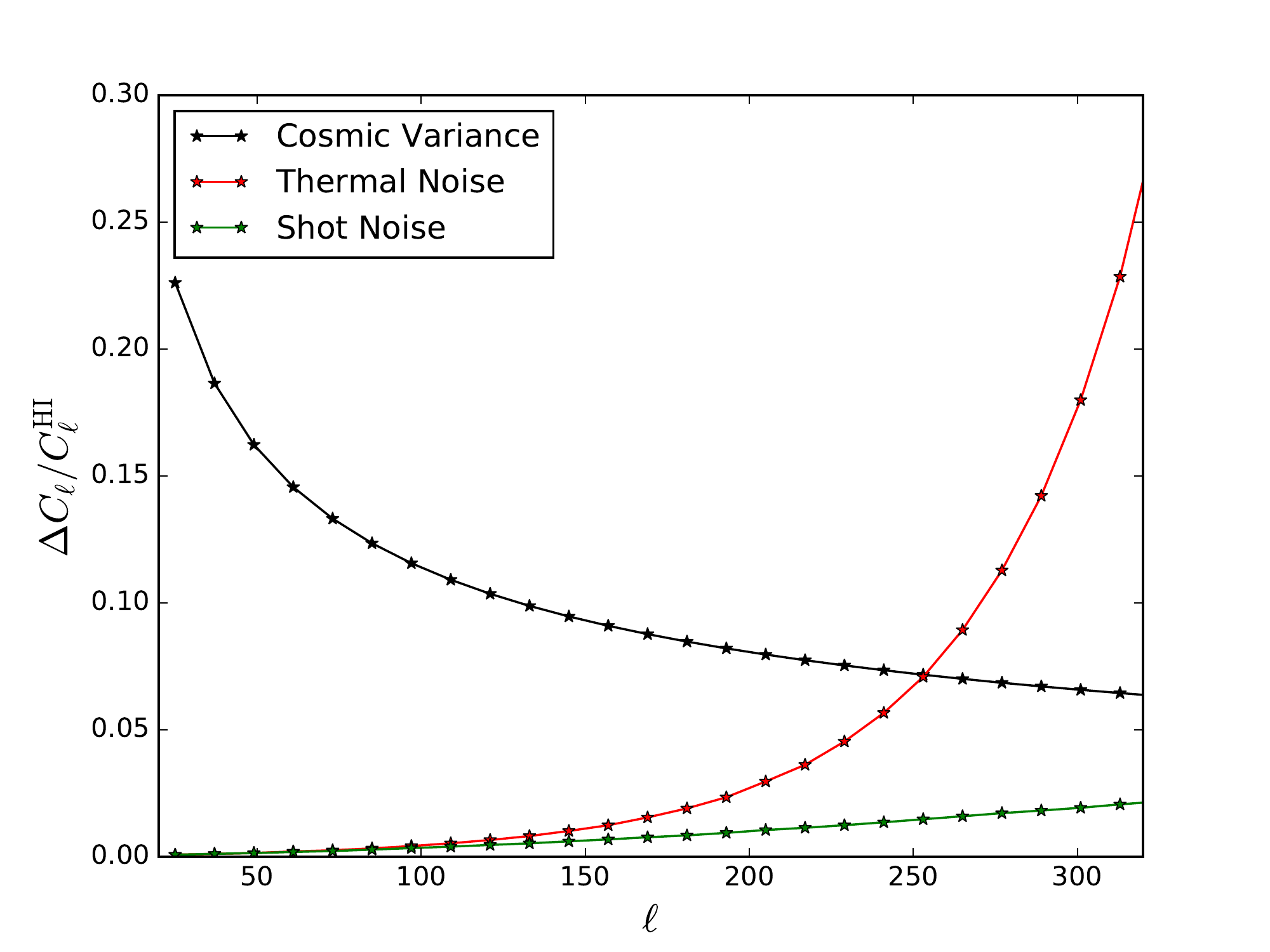}}
\subfigure{\includegraphics[width=0.49\textwidth]{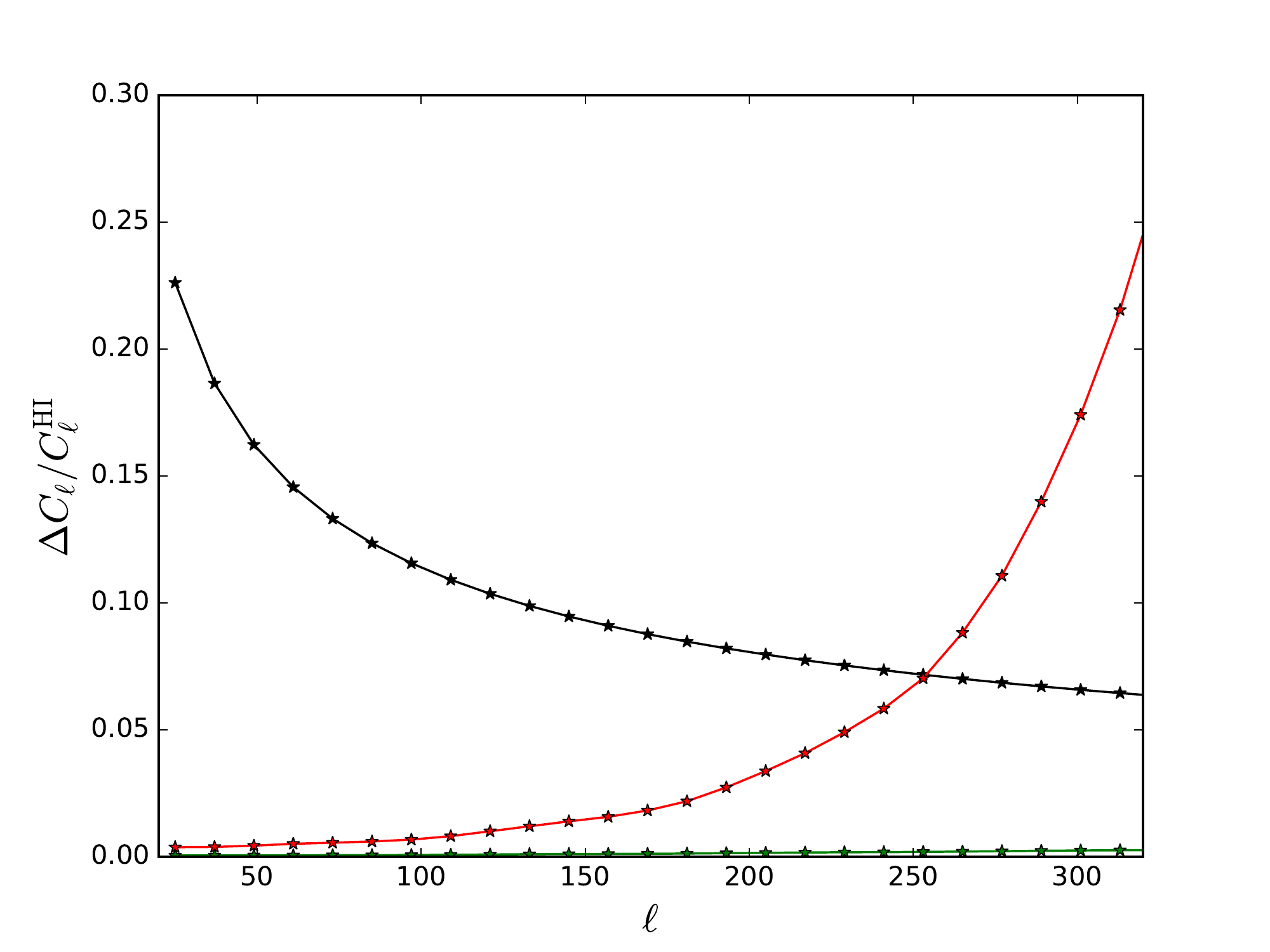}}
\caption{Fractional uncertainties in the recovered angular power spectrum for the BINGO experiment ($f_{\mathrm{sky}}=0.07$) for redshift $z = 0.13$ (\textit{left}) and $z = 0.48$ (\textit{right}). We plot the ratio between the different sources of uncertainty (cosmic variance, shot noise and thermal noise) and the HI angular power spectrum, $\Delta C_{\ell} / C_{\ell}^{\mathrm{HI}}$, as a function of multipole. For the cosmic variance we use the standard formula, $\Delta C_{\ell} / C_{\ell} = \sqrt{2 / (2 \ell + 1) f_{\mathrm{sky}} \Delta \ell}$, and for the thermal and shot noise we make $\Delta C_{\ell}$ to be equal to their angular power spectrum multiplied by the cosmic variance factor, $\sqrt{2 / (2 \ell + 1) f_{\mathrm{sky}} \Delta \ell}$. }
\label{fig:delta_cl}
\end{figure*}


\section{Methods}
\label{sec:methods}

In this section we briefly review two key analysis methods. The first is the GNILC method (Section~\ref{sec:gnilc}), which is used to extract the cosmological HI signal from the observed signal in the presence of foregrounds. The second is our methodology to estimate the cosmological parameter from the observed power spectra. We do this via a likelihood (Section~\ref{sec:likelihood}), which allows the full posterior probability distribution to be mapped out.

\subsection{Component Separation: GNILC}
\label{sec:gnilc}

To perform  any  HI IM  experiment we need to  subtract  the  astrophysical  contamination  that  will  be present  in  the  observed signal. It is therefore important to quantify the potential contaminating effects of foreground residuals. In  this  work,  we use the  Generalized  Needlet  Internal  Linear  Combination  (GNILC) method \citep{Remazeilles2011b,Olivari2016} as our component separation technique. The GNILC method is very versatile. It has been used in several contexts before. In \citet{Planck2016_3}, for instance, it has been successfully applied to \textit{Planck} data to disentangle two components of emission that suffer from spectral degeneracy, the cosmic infrared background (CIB) anisotropies and the Galactic thermal dust emission, and in \citet{Olivari2016}, it has been shown to work for HI IM experiments. The GNILC method uses both frequency and spatial information to separate the components of the observed data, which makes it to perform better than traditional frequency space Principal Component Analysis (PCA) and frequency-space parametric fitting.

The GNILC method can be divided into two main steps. First, using a prior on the HI and the thermal noise power spectra, the local ratio between the HI plus noise signal and the total signal is determined. This prior is given by the sum of the theoretically known HI power spectrum and the theoretically known noise power spectrum. Note that this prior is blind about the particular realization of the HI plus noise signal that is found in the observed (here, simulated) sky. This ratio is then used to perform a constrained PCA of the data to determine the effective dimension of the HI plus noise subspace. The second step is a multimensional ILC filter, which is done within the HI plus noise subspace found in the previous step. The ILC filter minimizes the foreground contamination that may still be present in the HI plus noise subspace. In the constrained PCA step, the number of principal components of the observation covariance matrix is estimated locally both in space and in angular scale by using a wavelet (needlet) decomposition of the observations. To make the selection of the principal components (foregrounds) of the observation covariance matrix, a statistical information criterion, the Akaike Information Criterion  \citep{Akaike1974}, is used. For more details of the formalism, see \citet{Olivari2016}.


\subsection{Likelihood}
\label{sec:likelihood}

Once we have a set of reconstructed HI plus noise maps, we can use these maps to calculate their angular power spectra and perform cosmological parameter estimation to study the large-scale properties of the Universe. We do this via a Bayesian likelihood analysis that allows the full posterior probability distribution to be mapped out. Assuming that the HI plus noise signal is approximately Gaussian, we can make a likelihood analysis using the calculated $C_{\ell}$'s, exactly as it is done for CMB. Since we are using the entire shape of the angular power spectrum, this increases the amount of information that can be obtained from an HI IM experiment compared to a simple BAO analysis \citep{Rassat2008}.  

Since we will be observing over some fraction of the sky, $f_{\mathrm{sky}}$, there is a non-negligible correlation between the different multipoles. In order to avoid calculating the covariance matrix between the different multipoles, which makes the calculation of the likelihood computationally demanding, we bin our angular power spectrum with a multipole resolution that respects the constraint $\Delta \ell > \pi / \gamma_{\mathrm{max}}$, where $\gamma_{\mathrm{max}}$ is the maximum angular resolution of the experiment. This makes the bins nearly independent simplifying the analysis. For BINGO, we have $\gamma_{\mathrm{max}} \approx 15^{\circ}$ so that we can choose $\Delta \ell = 12$, while for SKA, we have $\gamma_{\mathrm{max}} \approx 100^{\circ}$ so that we can choose $\Delta \ell = 4$. Binning  is not a problem as long as the spectrum is smooth on scales $\sim \Delta \ell$, which is the case for both BINGO and SKA.

We use the likelihood expression derived by \citet{Verde2003}:

\begin{align}
\label{like}
&-2 \ln \mathcal{L} = f_{\mathrm{sky}} \Delta \ell \sum_{i, \ell} (2 \ell + 1)  \\ \nonumber &\times \left[ \ln \left( \frac{C_{\ell}^{\mathrm{th}, i} + C_{\ell}^{\mathrm{s}, i} + N_{\ell}^i}{C_{\ell}^{\mathrm{ob}, i}} \right) + \frac{C_{\ell}^{\mathrm{ob}, i}}{C_{\ell}^{\mathrm{th}, i} + C_{\ell}^{\mathrm{s}, i} + N_{\ell}^i} - 1 \right],
\end{align} where $C_{\ell}^{\mathrm{th}, i}$ is the theoretical HI angular power spectrum for channel $i$ (see Eq.~\ref{c_ell_th}), $C_{\ell}^{\mathrm{s}, i}$ is the estimator of the shot noise power spectrum for channel $i$ (see Section~\ref{sec:shoot}), $N_{\ell}^i$ is the estimator of the thermal noise power spectrum for channel $i$, and $C_{\ell}^{\mathrm{ob}, i}$ is the power spectrum of the observed HI plus noise map for channel $i$.

The likelihood has three sources of uncertainties: cosmic variance, shot noise and thermal noise. The relative contribution of these uncertainties vary with multipole, $\ell$. In general, the shot and thermal noise also varies with frequency/redshift. For the BINGO simulations, however, we make, for simplicity, the thermal noise to be independent of frequency. For the much larger bandwidth of SKA-MID, on the other hand, we do account for this dependency (see Section~\ref{sec:ska}). As an example of the relative behaviour of these uncertainties, we plot the ratio of them to the HI angular power spectrum, $\Delta C_{\ell} / C_{\ell}^{\mathrm{HI}}$, as a function of $\ell$ for two values of redshifts for the BINGO experiment in Fig.~\ref{fig:delta_cl}. We see that the cosmic variance is the dominating source of uncertainty for large scales ($\ell \lesssim 250$) and that the thermal noise dominates at the small scales ($\ell \gtrsim 250$). At very low redshifts ($z \lesssim 0.15$) and large angular scales ($\ell \lesssim 100$) the shot noise has the same order of magnitude as the BINGO thermal noise, but they remain negligible compared to cosmic variance. In our analysis, we are ignoring any extra uncertainty on the measurement of the HI power spectra that arises from the foreground cleaning procedure. If we were to do this, the error bars on our power spectrum would increase. It should be noted, however, that, because GNILC projects the observed data into a subspace dominated by HI plus noise and performs an ILC analysis restricted to it, it makes, by construction, the foreground residual to be sub-dominant to the HI plus noise signal. The effects of the foregrounds, however, are analyzed in a different context -- the bias that may arise on the quantification of the HI power spectrum after a foreground cleaning procedure -- in Section \ref{sec:fore}. 

\begin{table}
  \footnotesize
   \caption{Cosmological parameters that we use in our analysis. For each parameter, we give the prior range and the value assumed in the baseline cosmology.}
    \label{tab:cosmology}
    \begin{center}
      \begin{tabular}{| l | c | c | l |}
        \hline
        Parameters &  Prior range & Baseline \\ \hline \hline
        $\Omega_b h^2$ & [0.005, 0.1] & 0.02224 \\ \hline
        $\Omega_c h^2$ & [0.001, 0.99] & 0.1198 \\ \hline
        $h$ & [0.2, 1.0] & 0.6727 \\ \hline
        $\tau$ & [0.01, 0.8] & 0.081 \\ \hline
        $n_s$ & [0.9, 1.1] & 0.9641 \\ \hline
        $\ln \, (10^{10} A_s)$ & [2.7, 4.0] & 3.096 \\ \hline
        $w$ & [$-$3.0, $-$0.33] & $-$1 \\ \hline
        $\sum m_{\nu}$ (eV) & [0, 5] & 0.06 \\ \hline
        $\Omega_{\mathrm{HI}} \, (\times 10^{4})$ & [4.0, 8.0] & 6.2 \\ \hline  
      \end{tabular}
    \end{center} 
\end{table}

To obtain constraints on the cosmological parameters, we then perform a Bayesian analysis using Eq.~$\eqref{like}$ as our likelihood. For this we use a modified version of the {\sc CosmoSIS} software \citep{Zuntz2015} and use the {\sc emcee} sampler \citep{Foreman2013} for our Monte-Carlo Markov Chain (MCMC). The {\sc emcee} sampler is a form of MCMC that uses an ensemble of `walkers' to explore the parameter space. Each walker chooses another at random and proposes along the line connecting the two of them using the Metropolis acceptance rule. For the number of chains, we use one for each walker, with the number of walkers, $n_{\mathrm{w}}$, being determined by the {\sc emcee} constraint: $n_{\mathrm{w}} \gtrsim 2 n_{\mathrm{p}} - 1$, where $ n_{\mathrm{p}}$ is the number of free parameters in the analysis. Our MCMC chains have at least 25000 sampling points and we stop them when all the parameter means and standard variations have become stable.

For our BAO wiggles analysis (see Section~\ref{sec:bao}), we assume that our likelihood respects a standard Gaussian distribution and that the different acoustic scales, which are a function of redshift, are independent from each other.


\section{Results}
\label{sec:results}

We now present the main results of our work, which are the forecasted parameter constraints for various scenarios and experimental setups.

\subsection{BINGO: Optimal Case}

\begin{table}
  \footnotesize
   \caption{Cosmological parameters for the $\Lambda$CDM model plus the HI density parameter. The \textit{Planck} constraints are from the 2015 release and include temperature and polarization data. BINGO constraints are from simulated data when considering the optimal case of HI signal plus thermal noise only (no systematics/foregrounds).}
    \label{tab:lcdm}
    \begin{center}
      \begin{tabular}{| l | c | c | c | c |}
        \hline
        Parameters & \textit{Planck} & \textit{Planck} + BINGO  \\ \hline \hline
        $\Omega_b h^2$ & $ 0.02224 \pm 0.00016$ & $ 0.02216 \pm 0.00013$ \\ \hline
        $\Omega_c h^2$ & $ 0.1198 \pm 0.0014$ & $ 0.1209 \pm 0.0008$  \\ \hline
        $h$ & $ 0.6727 \pm 0.0063$ & $ 0.6678 \pm 0.0037$  \\ \hline
        $\tau$ & $ 0.081 \pm 0.017$ & $ 0.075 \pm 0.016$ \\ \hline
        $n_s$  & $ 0.9641 \pm 0.0047$ & $ 0.9610 \pm 0.0040$ \\ \hline
        $\ln \, (10^{10} A_s)$ & $ 3.096 \pm 0.033$ & $ 3.087 \pm 0.032$ \\ \hline
        $\Omega_{\mathrm{HI}} \, (\times 10^{4})$ &\ldots & $ 6.31 \pm 0.12$\\ \hline
      \end{tabular}
    \end{center} 
\end{table}

Before considering the more realistic case of simulating BINGO data with foregrounds, we consider the optimal case with thermal and shot noise only, which means that we produce maps with HI signal  and thermal noise and use these maps in our cosmological analysis without performing any component separation. 

To make use of the likelihood Eq.~$\eqref{like}$, we need to calculate the angular power spectra of our maps. We do this by using the {\sc HEALPix} routine {\sc anafast} \citep{Gorski2005} and consider the multipole range $20 \leq \ell \leq 360$, which corresponds to angular scales $\approx 10^{\circ}$ to $\approx 0.5^{\circ}$. We do not use multipoles below 20 to constrain our cosmological parameters because BINGO has a limited sky coverage and does not probe the HI angular power spectrum at the very large angular scales. 

\begin{figure*}
\centering
\subfigure{\includegraphics[width=0.49\textwidth]{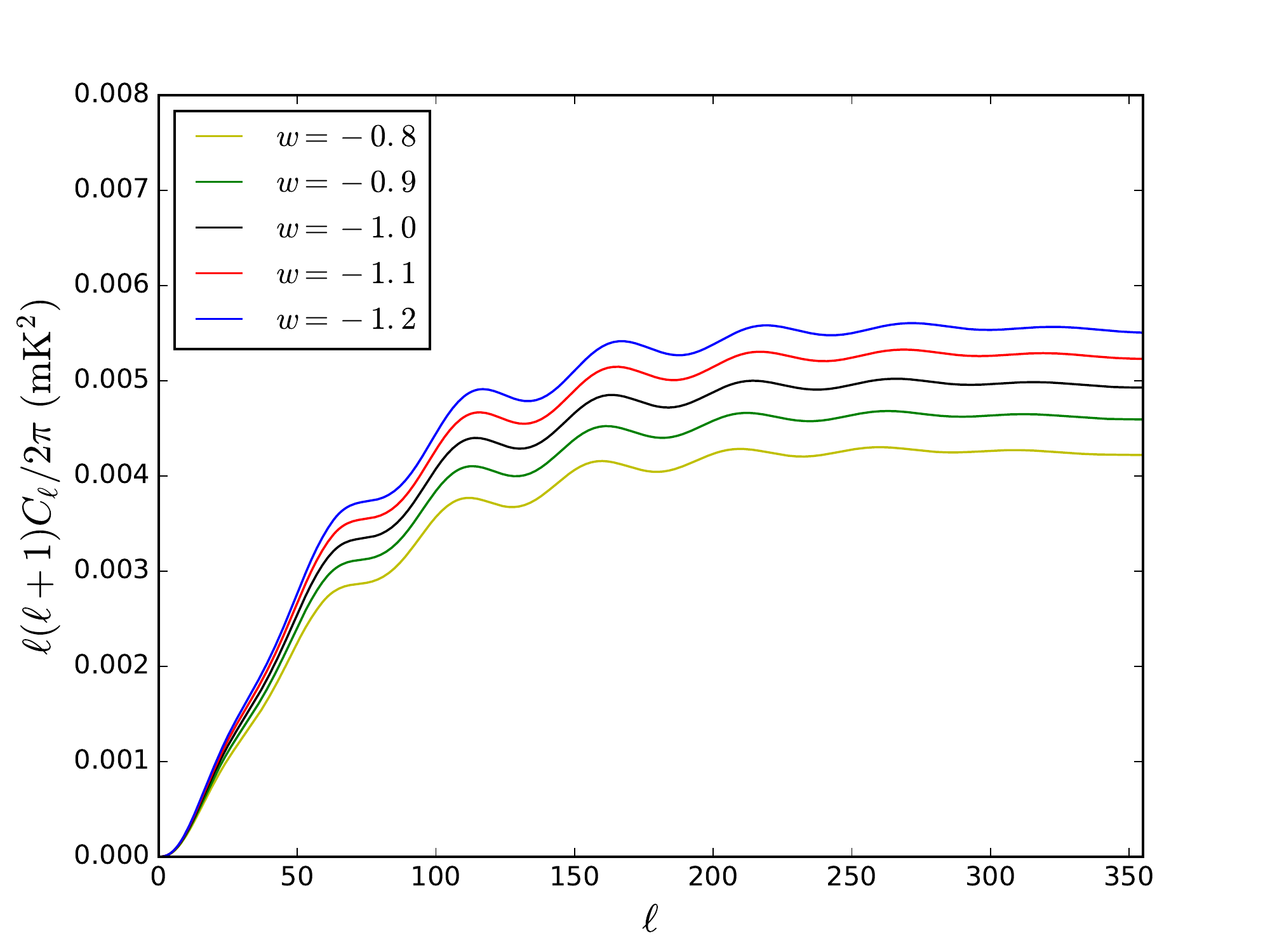}}
\subfigure{\includegraphics[width=0.49\textwidth]{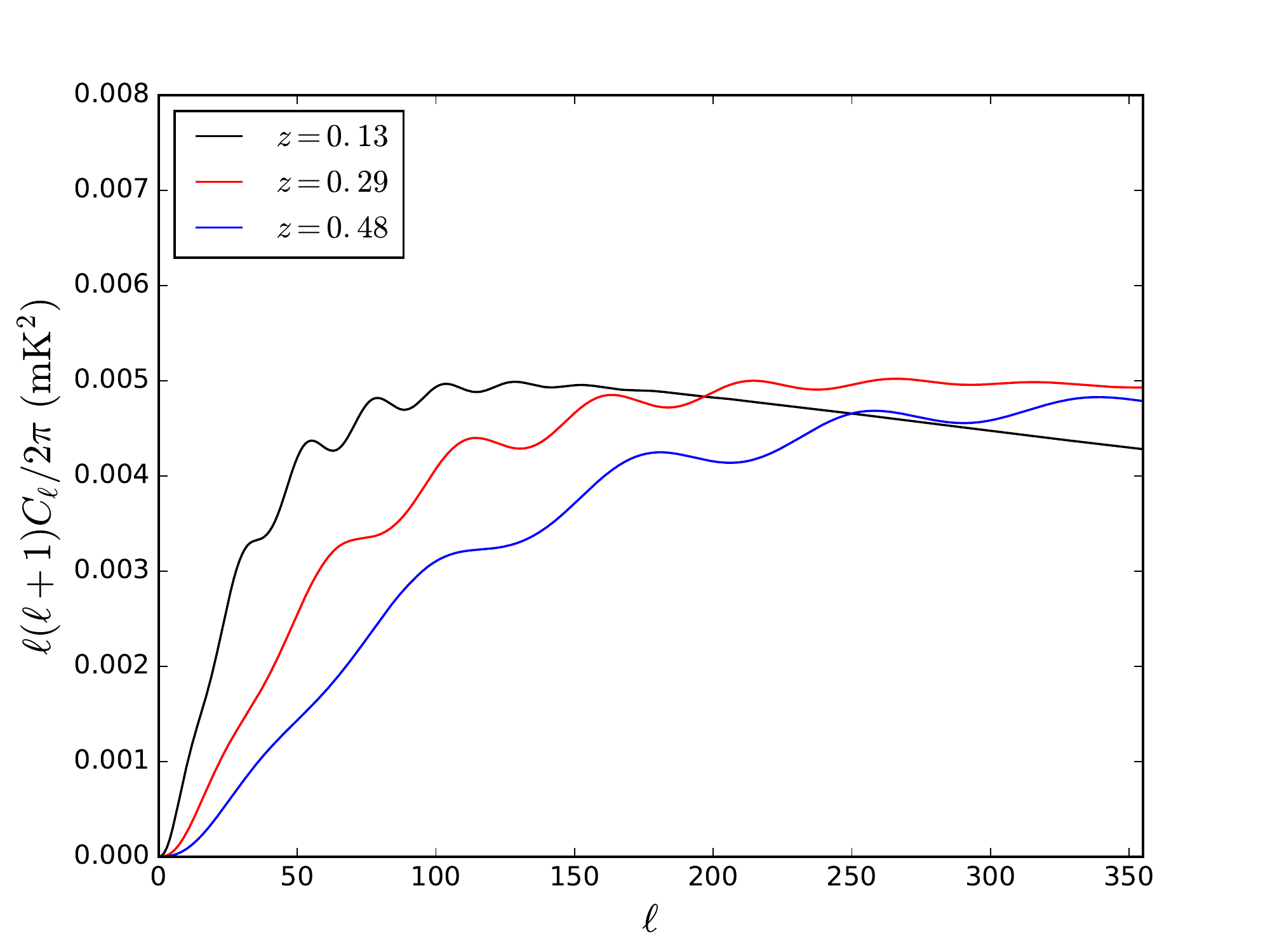}}
\caption{\textit{Left}: Dependence of the HI angular power spectrum, ${\ell (\ell + 1) C_{\ell} / 2 \pi}$, with the dark energy equation-of-state, $w$. For larger $w$, the HI angular power spectrum amplitude decreases and the BAO peaks move to larger scales. The opposite happens for smaller $w$: the amplitude increases and the BAO peaks move to smaller scales. We plot the HI angular power spectrum at $z = 0.3$ with $\Delta \nu$ = 7.5\,MHz. \textit{Right}: Dependence of the HI angular power spectrum on redshift for a fixed cosmology. For low redshifts ($z < 0.2$), the HI power, including the BAO peaks, is concentrated on large angular scales ($\ell < 100$). For higher redshifts, the HI power and the BAO peaks become more spread out in multipole space such that the HI signal contains now more power on small scales.}
\label{fig:cl_w}
\end{figure*}

To constrain the cosmological parameters, we combine the BINGO simulated data with the \textit{Planck} 2015 cosmological results. For this, we use the \textit{Planck} 2015 {\sc Plik lite} likelihood, which has the \textit{Planck} nuisance parameters marginalized as described in \citet{Planck2016_2}. We use the \textit{Planck} 2015 CMB temperature and polarization power spectra \citep{Planck2016_2}. We consider three cosmological models in our work: $\Lambda$CDM, $w$CDM, and $\Lambda$CDM with massive neutrinos. Table~\ref{tab:cosmology} lists the baseline cosmological parameters and ranges that we consider in our analysis, the results of which are presented in the following subsections.


\subsubsection{$\Lambda$CDM}

The current standard cosmological model is known as the $\Lambda$CDM model. This model corresponds to a spatially-flat, expanding Universe whose dynamics is governed by General Relativity and whose constituents are dominated at late times by cold dark matter (CDM) and a cosmological constant, $\Lambda$. The primordial seeds of structure formation are given by adiabatic fluctuations that respect a Gaussian distribution and that have an almost scale-invariant spectrum. This model is described by only six parameters: the baryon density today, $\Omega_b h^2$, the cold dark matter density today, $\Omega_c h^2$, the Hubble parameter, $h$, the amplitude of the primordial curvature perturbation, $A_s$, the scalar spectrum power-law index, $n_s$, and the optical depth due to reionization, $\tau$. In addition to these parameters we also have the parameters related to the HI signal. Initially we assume that the HI bias, $b_{\mathrm{HI}}$, is a constant equal to 1 so that the HI signal adds just one extra parameter to the set of six $\Lambda$CDM parameters: the $\Omega_{\mathrm{HI}}$ parameter (assumed for now to be independent of redshift), which determines the overall amplitude of the HI signal. In Section~\ref{sec:omg_hi} and \ref{sec:bias_hi}, we make the HI signal depend on more free parameters and analyze how these generalizations might affect the expected constraints on the cosmological parameters. 

When we combine \textit{Planck} with BINGO, we have a reduction in the uncertainties of the $\Lambda$CDM parameters compared to \textit{Planck} only. The most significant is the improvement by a factor of 2 on the uncertainty on the Hubble parameter, $h$, as can be seen in Table~\ref{tab:lcdm}. BINGO adds significantly more information at low redshift ($z \lesssim 0.5$) which helps break the degeneracies between $h$ and other parameters when considering primordial CMB data alone. We also find that the \textit{Planck}-plus-BINGO 6-parameter 1$\sigma$-uncertainty volume is smaller than that for \textit{Planck} alone by 80$\%$.  We note that the uncertainty we obtain here for $\Omega_{\mathrm{HI}}$ is an optimistic value since we are ignoring the calibration uncertainty that would to be present in the real data. Nevertheless, our results show that if the calibration uncertainty is not too large ($\lesssim 1\%$), BINGO has the potential to detect the HI signal overall amplitude with a good precision ($\Delta \Omega_{\mathrm{HI}} / \Omega_{\mathrm{HI}} \approx 2\%$).


\subsubsection{$w$CDM}
\label{subsec:wcdm}

Despite the observational success of the $\Lambda$CDM model, it is important to probe new physics beyond what is assumed in this model. One is related to the accelerated expansion of the Universe. The most straightforward candidate for dark energy is the cosmological constant $\Lambda$, which has a constant equation-of-state parameter $w = P / \rho = -1$. This model, however, exhibits some theoretical shortcomings such as the discrepancy between the value of the vacuum energy obtained through observations and the theoretically estimated value -- the so-called fine tuning problem of the cosmological constant. Most of the alternative models for dark energy that attempt to avoid the problems in the $\Lambda$CDM model make use of a dynamical field to describe the dark energy, e.g. quintessence \citep{Peccei1987, Ratra1988}. The simplest way to describe these models is by making the equation-of-state $w \neq -1$ but a free $w = w_0$ parameter. This phenomenological model is known as the $w$CDM model.

\begin{figure}
\centering
\includegraphics[width=0.51\textwidth]{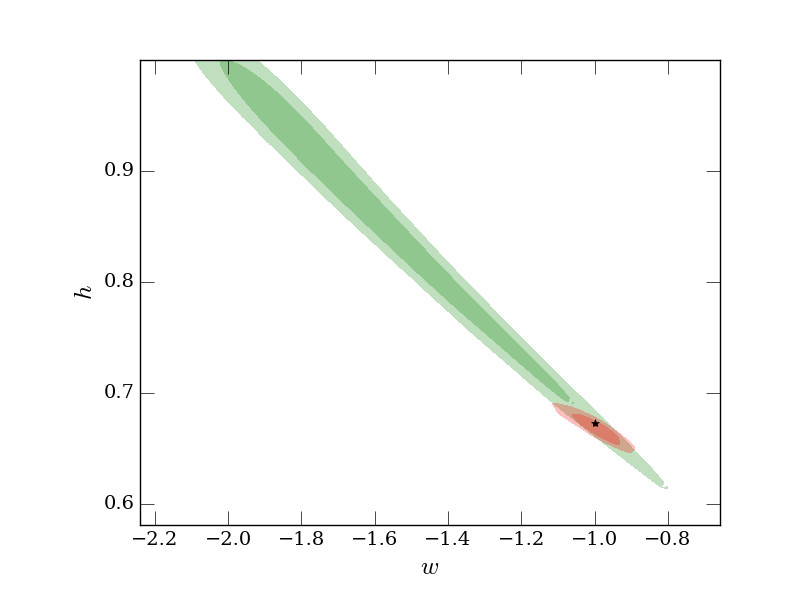}
\caption{2-dimensional likelihood contours for $h$ and $w$ for \textit{Planck} (\textit{green}) and \textit{Planck} plus BINGO for the optimal case without foregrounds (\textit{red}). The significant shrinking of the 2-dimensional contour that we obtain shows the power of BINGO data to break the degeneracy between $h$ and $w$ with CMB data alone. The black star corresponds to the baseline value of the parameters.}
\label{fig:2d_w_h}
\end{figure}

In Fig.~\ref{fig:cl_w}, we show how the dark energy equation-of-state changes the shape and amplitude of the HI angular power spectrum and how the HI angular power spectra evolve with redshift for a fixed cosmology. The evolution of the HI angular power spectrum with redshift depends mostly on the growth of the dark matter structure, which, for lower redshifts, depends, among other parameters, on $w$. For these reasons, the HI IM is able to break the degeneracy between $h$ and $w$, allowing the dark energy equation-of-state to be constrained much better than CMB data alone. This is clearly visible in Fig.~\ref{fig:2d_w_h}, where we plot the 2-dimensional contours for these two parameters for BINGO plus \textit{Planck} and for \textit{Planck} alone. The mean and 1$\sigma$ error (68$\%$ constraint) on $w$ that we obtain by combining \textit{Planck} with BINGO is
\begin{align}
\label{w_value}
&w = -1.00 \pm 0.04.
\end{align}
This means that a $4\%$ constraint could be achieved directly with a single IM experiment such as BINGO when combined with \textit{Planck} results.

We now compare the forecasted \textit{Planck} plus BINGO result with most recent combined constraints on $w$. \citet{Planck2016_1} showed that by using the \textit{Planck} temperature, polarization and lensing data, the 6dFGS, SDSS-MGS, and BOSS-LOWZ BAO measurements of $D_V / r_{\mathrm{drag}}$ \citep{Beutler2011, Ross2015, Anderson2014}, the CMASS-DR11 anisotropic BAO measurements of \citet{Anderson2014}, the JLA supernova sample \citep{Betoule2014}, and the Hubble parameter \citep{Efstathiou2014}, the $95\%$ constraint on the dark energy equation-of-state is $w = -1.019^{+0.075}_{-0.080}$. In comparison, we find that the $95\%$ constraint of the combination of BINGO and \textit{Planck} is $w = -0.996^{+0.072}_{-0.081}$, which has basically the same lower and upper limits than the previous result. More importantly, however, is the fact that IM surveys provide an alternative and independent cosmological probe, having a particular set of systematics (which will likely be different to other probes), which means that it has the potential to improve the confidence in dark energy results from current and future optical and near-infrared surveys.

\begin{figure}
\centering
\includegraphics[width=0.51\textwidth]{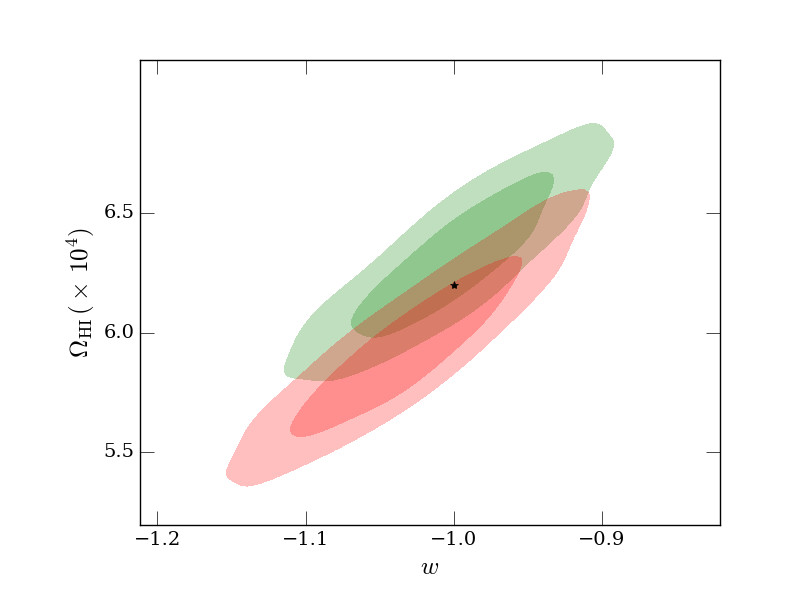}
    \caption{2-dimensional likelihood contours for $\Omega_{\mathrm{HI}}$ and $w$ for \textit{Planck} plus BINGO without foregrounds (\textit{green}) and \textit{Planck} plus BINGO with the foregrounds cleaned by the GNILC method (\textit{red}). The degeneracy between these two parameters is evident: both affect the overall amplitude of the HI angular power spectrum. Also note the bias on the parameters due to the slight underestimation of the HI power by the GNILC method. The black star corresponds to the baseline value of the parameters.}
    \label{fig:2d_w_omghi}
\end{figure}

\begin{figure}
  	\centering
    \includegraphics[width=0.51\textwidth]{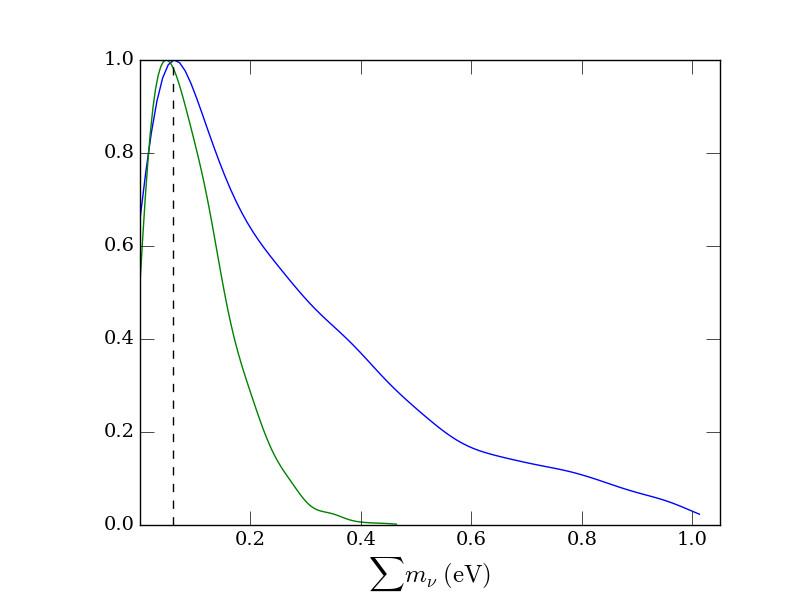}
    \caption{Marginalized posterior distributions for $\sum m_{\mathrm{\nu}}$ from \textit{Planck} (\textit{blue}) and \textit{Planck} plus BINGO without foregrounds (\textit{green}). Note the improvement that we obtain on $\sum m_{\mathrm{\nu}}$ by combining the BINGO simulated data with \textit{Planck} data. The dashed line is the baseline value $\sum m_{\mathrm{\nu}}=0.06$\,eV.}
    \label{fig:omg_nu}
\end{figure}


\begin{figure*}
  	\centering
\subfigure{\includegraphics[width=0.49\textwidth]{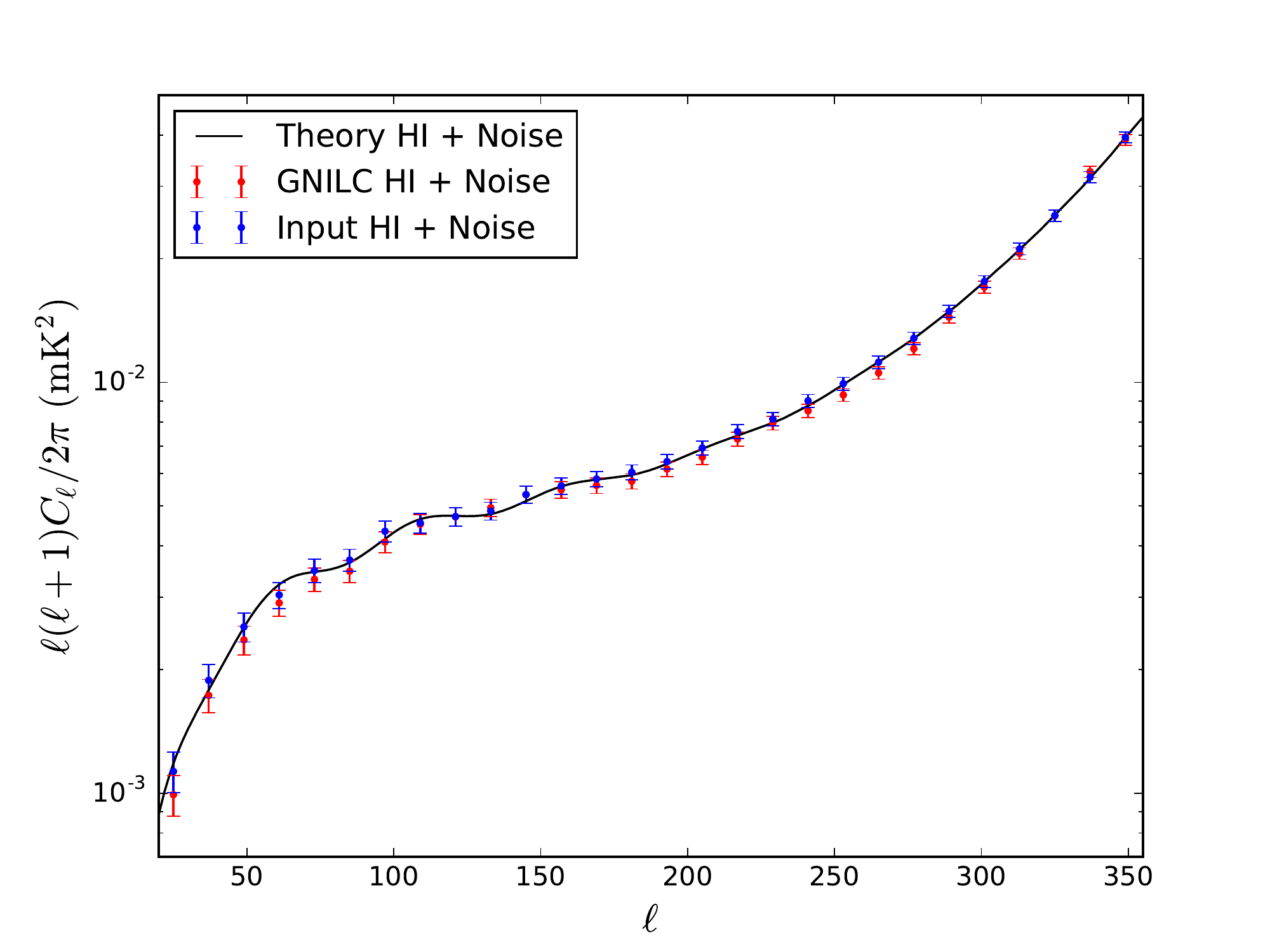}}
\subfigure{\includegraphics[width=0.49\textwidth]{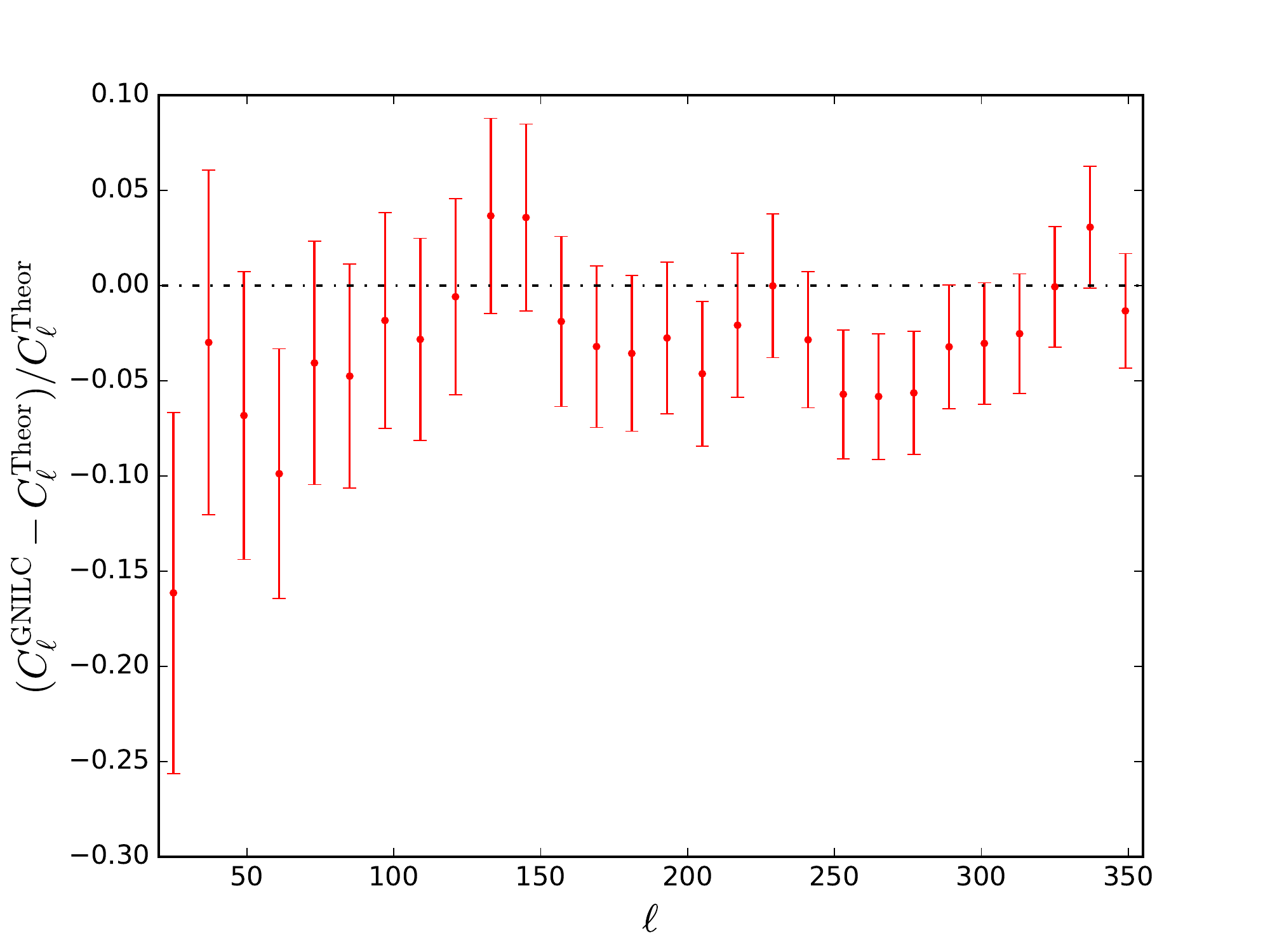}}
    \caption{\textit{Left}: Power spectra, ${\ell (\ell + 1) C_{\ell} / 2 \pi}$, at redshift $z = 0.3$, of the theoretical HI plus noise signal (\textit{black}), the HI plus noise signal as reconstructed by GNILC in the presence of foregrounds (\textit{red}), and the HI plus noise signal calculated from the input map without foregrounds (\textit{blue}). \textit{Right}: Fractional difference between the GNILC reconstructed power spectrum and the theoretical power spectrum, which, for a perfect foreground cleaning process, should be equal to zero. The error bars in both plots include cosmic variance, thermal noise, and shot noise uncertainties. The mean value of the red dots is $-3.1\%$.}
    \label{fig:gnilc}
\end{figure*}

We now make a few comments about our results. First, when using the HI angular power spectra, we find that what is mostly constraining the dark energy is their overall amplitude and shape, and not the BAO peaks. For instance, when we artificially remove BAO wiggles from the HI power spectra using the \citet{Eisenstein1998} approximation for the $P_c(k)$, we obtain $w = -0.98 \pm 0.05$, which is not significantly different from the value in Eq.~\eqref{w_value}. Second, it should be noted that the equation-of-state parameter has a degeneracy with $\Omega_{\mathrm{HI}}$ since both change the overall amplitude of the HI power spectrum. This degeneracy can be seen in Fig.~\ref{fig:2d_w_omghi}. This means that, by fixing $\Omega_{\mathrm{HI}}$ to its baseline value, we should obtain a better constraint on $w$, which is indeed what we find: $w = -1.02 \pm 0.02$, which is a factor of 2 better than the reference value of a $4\%$ uncertainty on $w$. 


\subsubsection{Massive Neutrinos}

The detection of solar and atmospheric neutrino oscillations proves that neutrinos have mass \citep{Ahmad2001, Fukuda1998}, with at least two species being non-relativistic today. The measurement of the absolute neutrino mass scale is of interest for both experimental particle physics and observational cosmology.

The main effect of total neutrino mass for CMB is around the first acoustic peak and is due to the early integrated Sachs-Wolfe (ISW) effect \citep{Lesgourgues2012}. The total neutrino mass also affects the angular-diameter distance to last scattering, and can be constrained through the angular scale of the CMB first acoustic peak. However, this effect is degenerate with $\Omega_{\Lambda}$ and $h$ in flat models. The use of CMB lensing or late-time measurements, like the BAO peaks of the HI power spectrum, can help in reducing this degeneracy. We therefore consider another extension of the $\Lambda$CDM model with massive neutrinos. To do this we let the sum of neutrino masses, $\sum m_{\mathrm{\nu}}$, to be a free parameter. Here, for simplicity, we assume two massless and one massive neutrino with a mass equal to $\sum m_{\nu}$.

Figure~\ref{fig:omg_nu} shows the improvement on the marginalized posterior distributions for $\sum m_{\mathrm{\nu}}$ by considering BINGO plus \textit{Planck} when compared with the \textit{Planck} case only. The 95$\%$ upper limit that we obtain on $\sum m_{\nu}$ from using \textit{Planck} only and combining \textit{Planck} with BINGO are
\begin{align}
\label{mass_nu_bingo}
&\sum m_{\nu} <  0.79 \; \mathrm{eV} \;\;\; [\mathrm{Planck}], \\ &\sum m_{\nu} < 0.24 \; \mathrm{eV} \;\;\; [\mathrm{Planck + BINGO}].
\end{align}
Again, we see that the IM data is having a large impact on the final uncertainty relative to CMB data alone.

We now compare the \textit{Planck} plus BINGO result with the current constraints on $\sum m_{\nu}$. The best 95$\%$ upper limit given by \citet{Planck2016_1}, which combines CMB, BAO, JLA, and $H_0$, is $\sum m_{\nu} <  0.23 \; \mathrm{eV}$, which is slightly stronger than the projected constraint from \textit{Planck} plus BINGO that we quoted above. However, we emphasize that we are not using JLA and $H_0$ data in the analysis above; when we do this we obtain the following constraint: $\sum m_{\nu} < 0.20 \; \mathrm{eV}$. The most important fact, however, is not the value itself, but the fact that this is an independent measurement of $\sum m_{\nu}$. As constraints (upper limits) improve towards the lower limit, independent measurements will be crucial in determining the true value of $\sum m_{\nu}$ if systematic errors are playing a significant role.


\subsection{BINGO: Effects of the Foregrounds}
\label{sec:fore}

To have a more realistic forecast of the BINGO experiment, we now investigate the impact of foregrounds emission on our projected constraints. We consider as our foregrounds the Galactic synchrotron, the Galactic free-free and the extragalactic point sources emissions, as described in Section~\ref{sec:foregrounds}. To clean this astrophysical contamination, we use the GNILC method (see Section~\ref{sec:gnilc}). 

\begin{table*}
  \footnotesize
   \caption{Effects of the foregrounds on the cosmological parameters of the $w$CDM model plus the HI density parameter. The \textit{Planck} data are the 2015 release and include temperature and polarization spectra. BINGO simulated data include by HI signal plus thermal noise without component separation (baseline) and by the GNILC reconstructed HI signal plus thermal noise (GNILC). We also include the difference between the \textit{Planck} + BINGO mean parameter value and the baseline value divided by the respective standard deviation, $\sigma$ (``parameter bias''). All parameter biases are within the interval $[-1.4, 1.4]\sigma$.} 
    \label{tab:wcdm}
    \begin{center}
      \begin{tabular}{| l | c | c | c | c | c | c |}
        \hline
        Parameters &  Baseline & \textit{Planck} + BINGO (baseline) & Parameter bias ($\sigma$) & \textit{Planck} + BINGO (GNILC) & Parameter bias ($\sigma$) \\ \hline \hline
        $\Omega_b h^2$ & 0.02224 & $ 0.02218 \pm 0.00014$ & $-0.43$ & $ 0.02218 \pm 0.00015$ & $-0.40$ \\ \hline
        $\Omega_c h^2$ & 0.1198 & $0.1205 \pm 0.0014$ & 0.50 & $ 0.1208 \pm 0.0013$ & $0.77$ \\ \hline
        $h$ & 0.673 & $0.668 \pm 0.009$ & $-0.55$ & $ 0.677 \pm 0.011$ & $0.36$ \\ \hline
        $\tau$ & 0.081 & $0.077 \pm 0.017$ & $-0.24$ & $ 0.069 \pm 0.016$ & $-0.75$ \\ \hline
        $n_s$ & 0.9641 & $0.9616 \pm 0.0047$ & $-0.53$ & $ 0.9582 \pm 0.0044$ & $-1.34$ \\ \hline
        $\ln \, (10^{10} A_s)$ & 3.096 & $3.089 \pm 0.033$ & $-0.21$ & $ 3.075 \pm 0.031$ & $-0.68$ \\ \hline
        $w$ & $-1.0$ & $ -1.00 \pm 0.04$ & $0.00$ & $ -1.03 \pm 0.05$ & $-0.60$ \\ \hline
        $\Omega_{\mathrm{HI}} \, (\times 10^{4})$ & 6.20 & $6.33 \pm 0.22$ & $0.59$ & $ 5.94 \pm 0.25$ & $-1.04$ \\ \hline
      \end{tabular}
    \end{center} 
\end{table*}

For the frequency channel corresponding to the redshift $z = 0.3$, we plot in Fig.~\ref{fig:gnilc} the theoretical HI plus noise power spectrum, the binned GNILC HI plus noise power spectrum, and the binned input HI plus noise power spectrum (calculated from the BINGO patch without any foreground) for the multipole range $20 \leq \ell \leq 360$ that we consider in our analysis. We also plot the normalized  difference of the input power spectrum and the GNILC reconstructed power spectrum, which, for a perfect foreground cleaning process, should be equal to zero. As can be seen in Fig.~\ref{fig:gnilc}, in cleaning the foregrounds contamination, the GNILC method loses some HI power, i.e., it underestimates the HI amplitude. The consequence of this is that there is a bias on the reconstruction of the HI angular power spectra. Using the 40 frequency channels that we consider in our cosmological analysis, we obtain an average normalized difference between the GNILC reconstructed power spectrum and the theoretical power spectrum of $-5.5\%$ (in Fig.~\ref{fig:gnilc} we show only one frequency channel, the middle one, which happens to have an average normalized difference of $\approx -3\%$).

We now consider the $w$CDM scenario and summarize our results at Table~\ref{tab:wcdm}, where we compare the cases with and without component separation. The component separation bias has mostly affected $n_s$ and $\Omega_{\mathrm{HI}}$ -- both parameters are more than 1$\sigma$ away from their baseline value. The bias on $\Omega_{\mathrm{HI}}$, for instance, can be visualized in Fig.~\ref{fig:2d_w_omghi}, where we plot the 2D-posterior for $w$ and $\Omega_{\mathrm{HI}}$. Although the biases on all the other parameter are inside their respective 1$\sigma$ error, their mean, as can be seen by comparing the parameter bias of the \textit{Planck} + BINGO (baseline) case with the parameter bias of the \textit{Planck} + BINGO (GNILC) case, have been affected by the foregrounds. For instance, $\Omega_{c} h^2$ and $\tau$ are almost three-quarters of a standard deviation away from their baseline value. 

We now make some comments about our simulations. First, our results are not completely general; if we include systematics for instance, we may obtain larger and therefore more significant biases on our cosmological parameters than the ones we quote here. Second, we are ignoring any extra uncertainty on the measurement of the HI power spectra that arises from the foreground cleaning procedure. If we do this, the error bars on our power spectrum will increase and the parameter bias will decrease (remember that the reason for the existence of a parameter bias due to an imperfect foreground cleaning procedure is the relatively high precision BINGO measurements of the HI power spectrum). Additionally, as can be seen in Fig. ~\ref{fig:gnilc}, because of the use of wavelets by GNILC, there is a non-negligible correlation between neighboring $\ell$ bins in the final estimation of the HI power spectrum. In a more general analysis, the quantification of this correlation should be present in the assumed likelihood. Finally, we have tested a single foreground model but the exact characteristics of the foreground sky are unknown. Given these facts, we should see our results as an indication of the level of accuracy that we may obtain on our cosmological parameters due to an imperfect foreground cleaning procedure. Nevertheless, the typical accuracy of a few \% (in the reconstruction of the HI power spectrum) that we see in our analyses is promising for IM experiments.


\subsection{Redshift Dependent HI Density}
\label{sec:omg_hi}

As BINGO and IM measurements in general will probe the HI emission across cosmic time, our simulation of the HI signal, to be more realistic, should consider an HI density parameter that evolves with redshift. So far, for simplicity, we have assumed the HI density parameter to be constant and used the value $\Omega_{\mathrm{HI}} = 6.2 \times 10^{-4}$ as measured by \citet{Switzer2013} using the Green Bank Telescope at $z \sim 0.8$. To consider the evolution of the HI density with redshift, we need to use a more complex model. We do this by considering the two-parameters model suggested by \citet{Crighton2015}:
\begin{equation}
\label{omg_hi}
\Omega_{\mathrm{HI}} = \Omega_{\mathrm{HI}}^0 (1 + z)^{\alpha},
\end{equation} 
where $\Omega_{\mathrm{HI}}^0 = 4 \times 10^{-4}$ and $\alpha = 0.6$.

We consider the $\Lambda$CDM and the $w$CDM scenarios, and, in addition to the cosmological parameters of these models, we allow $\Omega_{\mathrm{HI}}^0$ and $\alpha$ to be free parameters in our Bayesian analysis, increasing the complexity of our parameter space. For the $\Lambda$CDM cosmology, the results that we obtain by combining \textit{Planck} with BINGO (without component separation) for these two parameters are
\begin{align}
&\Omega_{\mathrm{HI}}^0 \, (\times 10^{4}) = 4.11 \pm 0.08, \\ \nonumber &\alpha = 0.57 \pm 0.03.
\end{align}
These results show us that BINGO has the potential to detect the evolution of $\Omega_{\mathrm{HI}}$ with redshift with a reasonable precision (2$\%$ for $\Omega_{\mathrm{HI}}^0$ and 5$\%$ for $\alpha$). For the $w$CDM cosmology, on the other hand, the results that we obtain by combining \textit{Planck} with BINGO (without component separation) are
\begin{align}
&w = -1.01 \pm 0.06 \\ \nonumber &\Omega_{\mathrm{HI}}^0 \, (\times 10^{4}) = 4.06 \pm 0.18, \\ \nonumber &\alpha = 0.57 \pm 0.04.
\end{align}

A more complex $\Omega_{\mathrm{HI}}$ leads to a larger uncertainty on $w$: the uncertainty on $w$ increases from $\Delta w = 0.04$ in the case of a constant $\Omega_{\mathrm{HI}}$ to $\Delta w = 0.06$ in the case of a two-parameters model for it. The reason for this increase is that we now have three quantities that change the overall amplitude of the HI angular power spectrum, $\Omega_{\mathrm{HI}}^0$, $\alpha$, and $w$, and these are degenerate with each other making  our determination of $w$ less precise. Also, because we now have one extra function of redshift in the HI temperature, Eq.~\ref{backtemp_1}, the ability of the BINGO experiment to discriminate between it and the growth factor is reduced. 

We make a final remark about the overall amplitude of the cosmological signal. As well as assuming a constant HI bias, we have ignored the overall amplitude calibration uncertainty associated with any real measurement. For a radio experiment this is likely to be in the range $\sim 1$--$10\%$ and thus any constraints relying on the absolute amplitude should take this additional source of uncertainty into account. However, parameters relying on the \textit{relative} amplitude of the HI power  as a function of redshift (for example) should be less affected since relative calibration is typically more accurate than absolute calibration.


\subsection{HI Bias}
\label{sec:bias_hi}

The main challenge for using the HI angular power spectrum to constrain cosmological parameters is the fact that we do not know how the HI bias evolves with redshift and scale \citep[see e.g.,][]{Padmanabhan2015, Penin2017}. This requires that we use some form of parametrization for them. As a case of study, we use polynomial models as this parametrization. For the redshift evolution, we use a three-parameter model,
\begin{equation} 
\label{bias_z}
 b_{\mathrm{HI}} (z) = b^z_{0} + b^z_{1} z + b^z_{2} z^2,
\end{equation} 
where $b^z_0 = 0.67$, $b^z_1 = 0.18$, and $b^z_2 = 0.05$, which fits the values for $b(z)$ from \citet{Bull2016}. For the dependency on wavenumber, we adapt the approach of \citet{Sarkar2016} to our redshift range, and use a three-parameter model
\begin{equation}
\label{bias_k} 
b_{\mathrm{HI}} (k) = b_0^k + b_1^k k + b_2^k k^2,
\end{equation}
where $b_0^k = 1.0$, $b_1^k = -1.1 \, \mathrm{Mpc}$, and $b_2^k = 0.4 \, \mathrm{Mpc}^{2}$. It should be noted that such models are not driven by data and therefore must be seen just as toy models.

In addition to the cosmological parameters of the $w$CDM model, we let the HI bias parameters (three parameters) and the HI density parameter (one parameter in this case) to be free in our Bayesian analysis. The results that we obtain for the case of the redshift-dependent HI bias using \textit{Planck} plus BINGO (without component separation) are
\begin{align}
\label{w_bias_1}
&w = -0.99 \pm 0.06, \\ \nonumber  &b^z_0 = 0.71 \pm 0.04, \\ \nonumber &b^z_1 =  0.15 \pm 0.11, \\ \nonumber &b^z_2 = 0.10 \pm 0.17, \\ \nonumber &\Omega_{\mathrm{HI}} \, (\times 10^{4}) = 6.11 \pm 0.40.
\end{align} 

For the case of the wavenumber-dependent HI bias using \textit{Planck} plus BINGO (without component separation) the results are
\begin{align}
\label{w_bias_2}
&w = -0.96 \pm 0.05, \\ \nonumber &b^k_0 = 1.06 \pm 0.09, \\ \nonumber &b^k_1 \, (\mathrm{in} \, \mathrm{Mpc})= -1.16 \pm 0.18, \\ \nonumber &b^k_2 \, (\mathrm{in} \, \mathrm{Mpc}^2) = 0.28 \pm 0.27, \, \\ \nonumber &\Omega_{\mathrm{HI}} \, (\times 10^{4}) = 6.22 \pm 0.50.
\end{align} 
We see that the HI bias affects the BINGO constraint on $w$. Its uncertainty, as expected, increases compared to the value we quote in Eq.~$\eqref{w_value}$.  The first parameter of the 3-parameters polynomial models has the most significant effect on the HI power spectrum; the other two, because of the small redshifts and relatively large scales being probed, do not significantly affect our observable, and hence they cannot be well constrained by BINGO. This, however, is not a general result. For experiments that are going to observe the HI signal at larger redshifts ($z > 1$), as, for instance, the SKA experiment, the higher order terms of the redshift expansion of the HI bias will have a more significant effect on the HI angular power spectrum.

\begin{table*}
  \footnotesize
   \caption{Summary of constraints on the dark energy equation-of-state $w$ (assuming the $w$CDM model) when using BINGO plus \textit{Planck} data, for  various different scenarios of complexity and analyses. The baseline value is $w=-1$.}
    \label{tab:wcdm_bingo}
    \begin{center}
      \begin{tabular}{| l | c |}
        \hline
        Scenario &  Constraint on $w$ (\textit{Planck} + BINGO)  \\ \hline \hline
        Power spectrum, no foregrounds and fixed $\Omega_{\mathrm{HI}}$ & $ -1.02 \pm 0.02$ \\ \hline
        Power spectrum and no foregrounds & $ -1.00 \pm 0.04$ \\ \hline
        Power spectrum with foregrounds cleaned by GNILC & $ -1.03 \pm 0.05$ \\ \hline
        Power spectrum, no foregrounds and redshift-dependent $\Omega_{\mathrm{HI}}$  & $ -1.01 \pm 0.06$ \\ \hline
        Power spectrum, no foregrounds and redshift-dependent $b_{\mathrm{HI}}$  & $ -0.99 \pm 0.06$ \\ \hline
        Power spectrum, no foregrounds and wavenumber-dependent  $b_{\mathrm{HI}}$  & $ -0.96 \pm 0.05$ \\ \hline
        BAO wiggles and no foregrounds & $-1.03 \pm 0.07$ \\ \hline
        BAO wiggles with foregrounds cleaned by GNILC & $-1.01 \pm 0.08$ \\ \hline
      \end{tabular}
    \end{center} 
\end{table*}

We now study the effect that we have on $w$ when we assume a model in our analysis that is different from the one that we use in our simulations. Here, we still simulate our HI maps with the 3-parameters polynomial models for the HI bias given by Eqs.~$\eqref{bias_z}$ and $\eqref{bias_k}$, but parametrize them with a power-law, $b_{\mathrm{HI}} = A (1 + B x)^{\alpha}$, where $x$ can either be $z$ or $k$ and $A$, $B$, and $\alpha$ are now the free parameters that we fit for. The result that we obtain in the case of the redshift-dependent HI bias using \textit{Planck} plus BINGO (no foregrounds) for the dark energy equation-of-state is
\begin{align}
&w = -0.98 \pm 0.05.
\end{align} 
In the case of the wavenumber-dependent HI bias, on the other hand, by using \textit{Planck} plus BINGO (no foregrounds), we obtain
\begin{align}
&w =  - 0.96 \pm 0.07.
\end{align}

The results above show that the values we quote for $w$ in Eqs. $\eqref{w_bias_1}$ and $\eqref{w_bias_2}$ are not strongly dependent on the assumed analytical formula of the HI bias. Of course, the above analysis is just a case of study and not a general procedure. Nevertheless, it indicates that as long as the HI bias has a mild dependency on redshift and scale, a simple analytical formula can take care of any potential bias on $w$, at least for sensitivity levels comparable to BINGO or other stage I/II experiments \citep{Albrecht2006,Bull2015}. 


\subsection{BAO}
\label{sec:bao}

Until now, we have been using the full shape of the HI angular power spectrum as our cosmological data. There is, however, another way to use the HI IM data to constrain the cosmological parameters, which is to isolate the baryon acoustic oscillations (BAO) from the HI angular power spectrum \citep{Battye2013}. With this method, the power spectrum of the two dimensional distribution of the HI signal is computed for each redshift bin and, assuming that the power spectrum of the HI is only biased relative to the underlying dark matter distribution by some overall scale-independent constant, the acoustic scale, which is a function of the cosmological parameters, can then be extracted.

In order to estimate the error on the measurement of the acoustic scale, $\ell_A$, we follow \citet{Blake2003} and fit a decaying sinusoidal function to the projected data, making the necessary adaptations from going from a 3-dimensional space to a 2-dimensional space. It is this a priori knowledge of the BAO wiggles that allows the acoustic scale to be measured accurately with relatively low signal-to-noise ratio. Once we have a measurement of $\ell_A$, we relate it to our cosmology by using the following relation,
\begin{equation}
\ell_A \approx \frac{2 \pi}{s} r(z),
\end{equation}
where $s$ is the size of the sound horizon and $r(z)$ is the comoving distance.

We use the same 40 frequency channels that we have used for the HI angular power spectra case. We assume that our likelihood is a standard Gaussian distribution and that each redshift bin is independent from each other. Ignoring the astrophysical foregrounds, we obtain the following constraint on $w$ by combining \textit{Planck} with the BAO as measured by BINGO,
\begin{equation}
\label{bao_w}
w = -1.03 \pm 0.07.
\end{equation}

The constraint on $w$ is worse by almost a factor of 2 when filtering the BAO wiggles compared to using the full shape of the HI angular power spectrum (Eq.~\eqref{w_value}). This result is expected because the full HI angular power spectra contains more statistical information than the BAO wiggles alone \citep{Rassat2008}. The use of the BAO wiggles, however, does have some advantages. The most important is that the BAOs do not depend on the overall amplitude of the HI angular power spectra, making them potentially more robust to foregrounds and the modelling of the HI bias.

We remark that our result is significantly better than the results presented by \cite{Battye2013} for two main reasons: first, the sky coverage of the BINGO experiment has increased due to the larger field-of-view resulting in a smaller cosmic variance, and second, the analysis of \cite{Battye2013} used a simplified binning (and shifting) of the BAO wiggles to a single redshift bin at $z=0.3$ resulting in some loss of information.

When considering foregrounds, we find smaller component separation biases on the cosmological parameters. The component separation bias on $n_s$ (the most affected parameter in the power spectrum case; see Section~\ref{sec:fore}), for instance, is now $-0.2\sigma$, which is significantly smaller than the value that we obtained by using the HI power spectra ($-1.3\sigma$ in this case). This is an example of where the BAO wiggles are potentially more robust than the power spectrum as a whole. For $w$, on the other hand, we obtain $w = -1.01 \pm 0.08$, which again is worse than the result that we obtain by using the power of the power spectrum in similar conditions (in the presence of foregrounds).    

We summarize all our constraints on $w$ from BINGO plus \textit{Planck} in Table~\ref{tab:wcdm_bingo}. The overall picture is that for a stage II experiment like BINGO $w$ can be constrained to a precision of better than $\approx 8\%$ using the BAO wiggles or $\approx 5\%$ using the power spectrum, as long as systematic errors can be controlled.



\begin{table*}
  \footnotesize
   \caption{Summary of constraints on the dark energy equation-of-state $w$ (assuming the $w$CDM model) and the sum of the neutrino masses $\sum m_{\nu}$ (assuming the $\Lambda$CDM with massive neutrinos model) when using SKA-MID plus \textit{Planck} data. The baseline values are $w=-1$ and $\sum m_{\nu} = 0.06\,$eV.}
    \label{tab:ska}
    \begin{center}
      \begin{tabular}{| l | c | c|}
        \hline
        Scenario &  $w$CDM & Massive neutrinos (95$\%$ upper limit) \\ \hline \hline
        SKA-MID band 1, no foregrounds & $w = -0.99 \pm 0.04$ & $\sum m_{\nu} < 0.11$\,eV \\ \hline
        SKA-MID band 2, no foregrounds & $w = -1.01 \pm 0.02$ & $\sum m_{\nu} < 0.08$\,eV \\ \hline
        SKA-MID band 2, no foregrounds and redshift-dependent $\Omega_{\mathrm{HI}}$  & $w = -1.01 \pm 0.03$ & $\sum m_{\nu} < 0.11$\,eV \\ \hline
        SKA-MID band 2, no foregrounds and redshift-dependent $b_{\mathrm{HI}}$  & $w = -1.02 \pm 0.04$ & $\sum m_{\nu} < 0.11$\,eV \\ \hline
       SKA-MID band 2, no foregrounds and wavenumber-dependent $b_{\mathrm{HI}}$  & $w = -0.99 \pm 0.03$ & $\sum m_{\nu} < 0.12$\,eV \\ \hline
      \end{tabular}
    \end{center} 
\end{table*}

\subsection{SKA}

As BINGO is considered a stage II experiment \citep{Albrecht2006,Bull2015}, its results should be comparable to current BAO measurements, such as WiggleZ \citep{Kazin2014} and BOSS \citep{Anderson2014}. HI IM experiments such as SKA \citep{Maartens2015}, which aims to study the formation and evolution of the first galaxies, dark matter, dark energy, and gravity, can perform better. Thus, to show the full potential of HI IM, we now consider briefly the SKA experiment using the auto-correlation approach, i.e., in total-power, single-dish mode.

We model the Phase 1 of the SKA-MID instrument as describe in Section~\ref{sec:ska}. In our likelihood analysis, as we have done for BINGO, we consider 40 frequency channels equally spaced in the frequency band. We bin our power spectrum with a multipole bin of $\Delta \ell = 4$ and our multipole range here depends on the frequency-dependent beam size, but for most of the cases it is given by $5 \leq \ell \leq 360$.

We consider two cosmological scenarios here: $w$CDM and $\Lambda$CDM with massive neutrinos. For simplicity, in what follows, no foregrounds, a constant and fixed $b_{\mathrm{HI}}$ and a constant but free $\Omega_{\mathrm{HI}}$ are assumed. For the $w$CDM scenario, the mean and 1$\sigma$ error on $w$ that we obtain by combining \textit{Planck} with SKA-MID band 1 are
\begin{align}
&w = -0.99 \pm 0.04 \;\;\; [Planck + \mathrm{SKA-MID \; band \; 1}],
\end{align}
and with SKA-MID band 2 are
\begin{align}
&w =  -1.01 \pm 0.02 \;\;\; [Planck + \mathrm{SKA-MID \; band \; 2}].
\end{align}

We see that band 2 of SKA-MID is better than band 1 to constrain the dark energy equation-of-state. The reason for this is two-fold: first, the redshift range of band 1, $z = [0.35, 3.0]$, is mostly outside the dark energy dominated epoch ($z \lesssim 1$) and does not go to very low redshift ($z \sim 0.1$) unlike band 2. Second, the beam size of band 1 is almost twice the size of band 2 so it cannot, for most of its frequency channels, explore the relevant high multipoles ($\ell \gtrsim 150$) of the angular power spectrum. 

For the $\Lambda$CDM with massive neutrinos case, the 95$\%$ upper limit that we obtain on $\sum m_{\nu}$ by combining \textit{Planck} with SKA-MID band 1 is
\begin{align}
&\sum m_{\nu} < 0.11 \; \mathrm{eV} \;\;\; [Planck + \mathrm{SKA-MID \; band \; 1}], 
\end{align}
and with SKA-MID band 2 is
\begin{align}
&\sum m_{\nu} < 0.08 \; \mathrm{eV} \;\;\; [Planck + \mathrm{SKA-MID \;  band \; 2}].
\end{align}

For the $\Lambda$CDM with massive neutrinos case, we see that both bands give similar results when combined with \textit{Planck} data. Both results are significantly better than the combination of BINGO with \textit{Planck} and of the current best constraint on $\sum m_{\nu}$ (upper 95$\%$ limit of 0.42 and 0.23\,eV, respectively). The above results are better than the current results even when more datasets are considered. \citet{Giusarma2016}, for instance, find an upper 95$\%$ limit of 0.18\,eV by adding the BOSS full shape of the power spectrum to the \textit{Planck} and BAO data and \citet{Vagnozzi2017} find  an upper 95$\%$ limit of 0.15\,eV by adding an external constraint on the optical depth to reionization, $\tau$. These results show that SKA-MID, because it can measure the overall shape of the HI power spectrum with good precision over a wide range of redshifts, has the potential to give results for $\sum m_{\nu}$ that approach the 95$\%$ lower limit ($\sum m_{\mu} > 0.06$\,eV) obtained from oscillation data \citep{Gonzalez-Garcia2016}. This particular result shows that the energy scale of the neutrino masses will be very soon in the reach of cosmology, particularly if different probes, such as redshift surveys and weak lensing surveys, are combined to constrain $\sum m_{\nu}$.

We now add a few extra complications to our SKA-MID band 2 analysis. First, we make $\Omega_{\mathrm{HI}}$ to be redshift dependent and then $b_{\mathrm{HI}}$ to be redshift and wavenumber dependent, using for this the models defined by Eqs. \ref{omg_hi}, \ref{bias_z}, and \ref{bias_k}. For the $w$CDM scenario, the new results are

\begin{align}
&w = -1.01 \pm 0.03 \;\;\; [\mathrm{Redshift-dependent}\;\Omega_{\mathrm{HI}}], \\ \nonumber
&w = - 1.02 \pm 0.04 \;\;\; [\mathrm{Redshift-dependent}\;b_{\mathrm{HI}}], \\ \nonumber
&w = - 0.99 \pm 0.03 \;\;\; [\mathrm{Wavenumber-dependent}\;b_{\mathrm{HI}}].
\end{align} For $\Lambda$CDM with massive neutrinos scenario, the new results are

\begin{align}
&\sum m_{\nu} < 0.11 \; \mathrm{eV} \;\;\; [\mathrm{Redshift-dependent}\;\Omega_{\mathrm{HI}}], \\ \nonumber
&\sum m_{\nu} < 0.11 \; \mathrm{eV} \;\;\; [\mathrm{Redshift-dependent}\;b_{\mathrm{HI}}], \\ \nonumber
&\sum m_{\nu} < 0.12 \; \mathrm{eV} \;\;\; [\mathrm{Wavenumber-dependent}\;b_{\mathrm{HI}}].
\end{align} The overall picture is then that SKA-MID band 2 can constrain $w$ to a precision of better than $\approx 4\%$ and can obtain a upper limit (95$\%$) of $\lesssim 0.12$ for $\sum m_{\nu}$ using the power spectrum, as long as systematic and foregrounds errors can be controlled. Note that more conservative analyses than ours put an upper limit (95$\%$) of $\lesssim 0.2$\,eV for $\sum m_{\nu}$ \citep[e.g.,][]{Villaescusa2015}.

We summarize our baseline results for SKA phase-1 operating in auto-correlation mode in Table~\ref{tab:ska}.


\section{Conclusions and Discussion}
\label{sec:conclusion}

The main goal of this work was to prove the feasibility of using the HI angular power spectrum to constrain the cosmological parameters, and in particular, the dark energy equation-of-state $w$. To achieve this, we first made an idealized simulation of the BINGO experiment, in which we assumed that there was no foregrounds and that the HI amplitude could be parametrized by a single parameter, and then made several complications to this simple simulation to see how robust our baseline results were. A summary of the main results on the constraints on $w$ are presented in Table~\ref{tab:wcdm_bingo}.

In the ideal case, with no systematics or foregrounds, we found that the combination of BINGO and \textit{Planck} 2015 data can constrain $w$ to a precision of $4\%$ when fitting for the power spectrum. This can be considered as the optimal baseline result for a stage II experiment like BINGO. 


The first complication was then to consider the presence of foregrounds emissions in our data. To remove these foregrounds from the observed data we used the GNILC algorithm \citep{Olivari2016}. Although a small ($\approx 6\,\%$) loss of HI power was observed, it was not large enough to significantly bias our cosmological parameters. Only two of them, $n_s$ and $\Omega_{\mathrm{HI}}$, suffered a bias of more than 1$\sigma$. The uncertainty on $w$ was increased to $5\%$.

The second complication was to let the HI density parameter evolve with redshift. For this, we chose a two-parameter model (see Eq.~$\eqref{omg_hi}$) for $\Omega_{\mathrm{HI}}$, which led to a larger uncertainty on $w$ ($\Delta w=0.06$). The reason for this is that now three quantities change the overall amplitude of the HI angular power spectrum, $\Omega^0_{\mathrm{HI}}$ , $\alpha$, and $w$, and these are degenerate with each other making our determination of $w$ less precise. 

The third complication was to consider a redshift-dependent and a wavenumber-dependent HI bias, using 3-parameters polynomial models (Eqs. \eqref{bias_z} and \eqref{bias_k}).  For the redshift-dependent case, we obtained a uncertainty of $6\%$ for $w$, while for the wavenumber-dependent case, we obtained a uncertainty of $5\%$ for $w$. We also found that even if we choose the wrong parametrization of the HI bias, we can still measure the dark energy equation-of-state without a strong bias or increase in its uncertainty. This situation, however, could be different if we go to higher redshifts or probe smaller or larger scales than BINGO.

We also considered the use of the BAO wiggles alone as our BINGO cosmological data, rather than using the full power spectrum. As expected, it gave us worse constraints than the use of the HI angular power spectra by approximately a factor of 2. The use of the BAO wiggles, however, showed to be more robust to foregrounds since it does not depend on the overall amplitude of the HI angular power spectra. Yet, even in the presence of extra nuisance parameters for the HI bias and amplitude, the power spectrum appears to still be potentially more powerful than the BAO wiggles alone. Clearly, both types of analyses should be made, providing a quasi-independent measurement of $w$ with potentially different limiting systematics.

To show the full potential of HI IM, we also considered briefly the Phase 1 of the SKA-MID experiment in our work. Assuming only thermal noise, we found that its band 2 can give us a significantly better constraint on $w$ ($\Delta w = 0.02$) than BINGO, while its band 1 can give us a similar constraint than BINGO ($\Delta w = 0.04$). For the $\Lambda$CDM with massive neutrinos case, however, we found that both bands when combined with \textit{Planck} give a better constraint on $\sum m_{\nu}$ than the one we obtained by combining BINGO with \textit{Planck}. We have also found that even in the presence of extra nuisance parameters, the results obtained by combining SKA-MID band 2 with \textit{Planck}  -- $\Delta w \lesssim 0.04$ and an upper limit (95$\%$) on $\sum m_{\nu}$ of $\lesssim 0.12$ -- are significantly better than the current results, which, it is worth to say, make use of CMB, BAO, supernovae and $H_0$ data while in our work we have combined SKA only with \textit{Planck}.

The main shortcoming of our analysis, as we mentioned before, is our simplification of the BINGO and the SKA experimental setups. We have ignored several systematics that may be present in the real data, such as $1/f$ noise, atmospheric noise, standing waves, real beams and calibration errors. These systematics will potentially make the HI signal reconstruction process less precise than the 6$\%$ level that we have obtained in this work for the case of BINGO data in the presence of foregrounds. However, as has been stated before, as cosmological probes become more precise, the \textit{accuracy} of the results in the presence of systematic errors will become the limiting factor. Therefore, independent probes, using different techniques with different systematics, will be essential for understanding discrepancies between datasets. We also used a simple model for the HI signal. We have ignored, for example, the correlation between different redshift bins and the redshift space distortion contribution to the HI angular power spectrum. These effects can be important and should be considered for future analyses. This is particularly true if we want to use HI IM to constrain the combination of growth rate with the normalization of the matter power spectrum, $\sigma_8 f(z)$, which can be used to constrain modified gravity models \citep{Hall2013, Santos2015, Bull2016}. 

Given the results presented here and the development of IM surveys in general,  we can expect IM to become a major tool for cosmology in the next few years, particularly with large facilities coming online over the next decade such as the SKA.


\section*{Acknowledgments}
LCO acknowledges funding from CNPq, Conselho Nacional de Desenvolvimento Cient\'{i}fico e Tecnol\'{o}gico - Brazil. CD and MR acknowledge support from an ERC Starting (Consolidator) Grant (no.~307209). CD and RAB acknowledge support from an STFC Consolidated Grant (ST/P000649/1). AAC acknowledges FAPESP, S\~{a}o Paulo Research Foundation, for financial support under grant number 2016/04797-9.
  
\bibliography{lucas_refs}
\bsp

\label{lastpage}

\end{document}